\documentclass[12pt]{iopart}

 \usepackage{color}
 \usepackage{epsfig}
 \usepackage{epstopdf}
 \usepackage{iopams}
 \usepackage{cite} 
 
 \expandafter\let\csname equation*\endcsname\relax
 \expandafter\let\csname endequation*\endcsname\relax

\usepackage{graphicx}
\usepackage{dcolumn}
\usepackage{bm}
\usepackage{braket}
\usepackage{MnSymbol}
\usepackage{scalerel,stackengine}
\newcommand\reallywidehat[1]{%
\savestack{\tmpbox}{\stretchto{%
  \scaleto{%
    \scalerel*[\widthof{\ensuremath{#1}}]{\kern-.6pt\bigwedge\kern-.6pt}%
    {\rule[-\textheight/2]{1ex}{\textheight}}
  }{\textheight}%
}{0.5ex}}%
\stackon[1pt]{#1}{\tmpbox}%
}

\newcommand{\pb}{\mathbf p}
\newcommand{\pbn}[1]{\mathbf p^{(#1)}}
\newcommand{\pn}[1]{p^{(#1)}}
\newcommand{\pbnt}[1]{\tilde{\mathbf p}^{(#1)}}

\begin{document}

\title[Stochastic thermodynamics of self-oscillations: the electron shuttle]{Stochastic thermodynamics of self-oscillations: the electron shuttle}

\author{Christopher W. W\"achtler${}^{1}$, Philipp Strasberg${}^{2,3}$, Sabine H. L. Klapp${}^{1}$, Gernot Schaller${}^{1}$ and Christopher Jarzynski${}^{4}$}

\address{$^1$ Institute of Theoretical Physics, Secr. EW 7-1, Technical University Berlin, Hardenbergstr. 36, D-10623 Berlin, Germany}
\address{$^2$ Complex Systems and Statistical Mechanics, Physics and Materials Science, University of Luxembourg, L-1511 Luxembourg, Luxembourg}
\address{$^3$ F\' isica Te\` orica: Informaci\' o i Fen\` omens Qu\` antics, Departament de F\' isica, Universitat Aut\` onoma de Barcelona, ES-08193 Bellaterra (Barcelona), Spain}
\address{$^4$ Institute for Physical Science and Technology, University of Maryland, College Park, MD 20742 USA}
\ead{christopher.w.waechtler@campus.tu-berlin.de}

\begin{abstract} 
Self-oscillation is a phenomenon studied across many scientific disciplines, including the engineering of efficient heat engines and electric generators. We investigate the single electron shuttle, a model nano-scale system that exhibits a spontaneous transition towards self-oscillation, from a thermodynamic perspective. We analyze the model at three different levels of description: The fully stochastic level based on Fokker-Planck and Langevin equations, the mean-field level, and a perturbative solution to the Fokker-Planck equation that works particularly well for small oscillation amplitudes. We provide consistent derivations of the laws of thermodynamics for this model system at each of these levels. 
At the mean-field level, an abrupt transition to self-oscillation arises from a Hopf bifurcation of the deterministic equations of motion. 
At the stochastic level, this transition is smeared out by noise, but vestiges of the bifurcation remain visible in the stationary probability density.
At all levels of description, the transition towards self-oscillation is reflected in thermodynamic quantities such as heat flow, work and entropy production rate.
Our analysis provides a comprehensive picture of a nano-scale self-oscillating system, with stochastic and deterministic models linked by a unifying thermodynamic perspective.
\end{abstract}

\maketitle

\section{\label{sec:Introduction}Introduction}
Self-oscillation has been described as "the generation and maintenance of a periodic motion by a source of power that lacks a corresponding periodicity" \cite{JenkinsPR2013}. As opposed to resonant systems, in which the driving source is modulated externally, the energy required to sustain self-oscillations is supplied by a 
constant source. The phenomenon is familiar from everyday life, e.g. the human voice and the sound 
of a violin string. Autonomous oscillations appear in a wide range of biological systems and chemical
and biochemical processes \cite{VanDerPolVanDerMarkJoS1928, NovakTysonNature2008, FieldNoyesJoCP1974} controlling, e.g., the beating of the 
heart, circadian cycles in body temperature or the Belousov-Zhabotinsky reaction. By converting direct current into stable oscillations, self-oscillatory systems provide a useful transduction mechanism for 
the design of autonomous motors and heat engines.

One particularly interesting system exhibiting self-oscillation is the electron shuttle, first proposed by 
Gorelik \textit{et al.} \cite{GorelikEtAlPRL1998}, where the mechanical oscillation of a metallic grain is achieved by sequential electron tunnelling between the grain and two connecting leads. This coupled system of mechanical and electronic degrees of freedom 
has drawn considerable theoretical and experimental attention since its original proposal. Theoretical 
descriptions of the system range from full quantum mechanical models of the coherent 
dynamics \cite{BoeseSchoellerEPL2001, ArmourMacKinnonPRB2002, MccarthyEtAlPRB2003, NovotnyEtAlPRL2003, NovotnyEtAlPRL2004, DonariniEtAlNJP2005,UtamiEtAlPRB2006} to semiclassical 
\cite{FedoretsEtAlEPL2002, UtamiEtAlPRB2006, NoceraEtAlPRB2011} and completely classical descriptions 
\cite{GorelikEtAlPRL1998, IsacssonEtAlPhysicaB1998, WeissZwergerEPL1999, NordEtAlPRB2002}. The electron shuttle has been experimentally realized by a 
vibrational fullerene molecule \cite{ParkEtAlNature2000}, gold grains 
\cite{MoskalenkoEtAlPRB2009, MoskalenkoEtAlNanotechnology2009, KonigWeigAPL2012} as well as nanopillars \cite{ScheibleBlickAPL2004} as molecular junctions between two leads. Also macroscopic electron shuttles, consisting of a pendulum between two capacitor plates, have 
been investigated \cite{KimEtAlAPL2015}. Reviews on 
the electron shuttle can be found in Refs.~\cite{JoachimEtAlNature2000, ShekhterEtAlJoP2003, GalperinEtAlJoP2007, GalperinEtAlScience2008, ShekhterEtAlNano2013, LaiEtAlFoP2015}.

Classical self-oscillating systems have been analyzed using the tools of deterministic non-linear dynamics \cite{JenkinsPR2013,StrogatzBook2018}.
For the electron shuttle in particular,
a sharp transition from stationarity to self-oscillation arises due to a Hopf bifurcation as the voltage difference between the leads crosses a threshold value \cite{UtamiEtAlPRB2006}.
While the dynamical description is well understood, the electron shuttle has not yet been thoroughly investigated from a thermodynamic perspective. Our aim in this paper is to provide such a perspective and lay out the groundwork for further thermodynamic analysis of the electron shuttle as a paradigmatic isothermal engine that converts direct electric current into periodic mechanical motion \cite{WaechtlerEtAlArXiv2019}. Because the electron shuttle is a nanoscale device, fluctuations play a central role in our analysis, in contrast with deterministic classical models.

In our analysis we will apply the tools of stochastic thermodynamics, a framework that formulates the laws of thermodynamics at the single-trajectory level and is particularly useful for investigating the thermodynamic behavior of nanoscale systems.
As described in review articles and monographs~\cite{EspositoHarbolaMukamelRMP2009, SekimotoBook2010, CampisiHaenggiTalknerRMP2011, JarzynskiAnnuRevCondMat2011, SeifertRPP2012, SchallerBook2014, VandenBroeckEspositoPhysA2015}, stochastic thermodynamics has been applied to a wide range of topics, including far-from-equilibrium fluctuation theorems, the operation of biomolecular machines, feedback control of nanoscale systems, the thermodynamic arrow of time, and the thermodynamic implications of information processing.

In recent years a number of models of non-autonomous stochastic heat engines have been proposed and investigated within the stochastic thermodynamic framework \cite{SchmiedlSeifertEPL2007, BlickleBechingerNature2011, RanaEtAlPRE2014, HolubecJSM2014, MartinezNature2016, RestrepoEtAlNJP2018}.
In these models, externally applied time-periodic driving leads to the conversion of thermal fluctuations into work. 
By contrast, autonomous nano-scale engines are characterized by the absence of an externally imposed cycle.
A variety of such autonomous engines have recently drawn both theoretical and experimental attention. First, thermoelectric devices, which use the interplay 
of thermal and chemical gradients to perform useful tasks, were proposed \cite{SegalPRL2008, EspositoEtAlEPL2009, SanchezButtikerPRB2011, StrasbergEtAlPRL2013, SothmannEtAlNano2014, BenentiEtAlPhysRep2017} and experimentally realized using quantum dot (QD) structures \cite{FeshchenkoEtAlPRB2014, HartmannEtAlPRL2015, ThierschmannEtAllNano2015}. Second, stochastic self-oscillatory engines were analyzed, including a Brownian gyrator \cite{FilligerReimannPRL2007, ChianEtAlPRE2017}, a rotor engine \cite{RouletEtAlPRE2017, FogedbyImparatoEPL2018}, a heat engine based on  Josephson junctions \cite{MarchegianiEtAlPRA2016}, solar cells \cite{AlickiEtAlJPA2015, AlickiEtAlAP2017, AlickiEntropy2016}  or mechanical resonators \cite{SerraGarciaEtAlPRL2016, ChianEtAlPRE2017}, and an experimental realization of the Feynman's ratchet-and-pawl mechanism \cite{Bang_etalNJP2018}.
While particular thermodynamic aspects such as nonequilibrium hot electron transport \cite{KochEtAlPRB2004}, subresonance inelastic electronic transport \cite{GalperinEtAlPRB2009, RomanoEtAlPRB2010} and tip-induced cooling \cite{SchulzeEtAlPRL2008} have been investigated, a systematic thermodynamic description of a nano-scale self-oscillating system, such as the one we provide for the electron shuttle, is still missing.

We investigate the electron shuttle at three different levels of description: the fully stochastic level modeled by a Fokker-Planck equation (FPE) and the equivalent Langevin equation, a mean-field (MF) model described as a deterministic dynamical system, and an intermediate perturbative model based on multiple scale (MS) perturbation theory, containing both deterministic and stochastic elements.
We study the dynamics and obtain statements of the first and second laws of thermodynamics at all three levels of description.
In doing so, we draw a direct line between our stochastic thermodynamic model of the electron shuttle and the nonlinear dynamic model of Refs. \cite{GorelikEtAlPRL1998, IsacssonEtAlPhysicaB1998}.
We find that the abrupt onset of self-oscillatory behavior observed at the deterministic level, appears at the stochastic level as a smoothed but nevertheless discernible transition from stationarity to self-oscillation.
At all three levels of description, this transition is reflected in thermodynamic quantities such as the rates of heat flow and entropy production.

\textit{Outline}: The article starts with a short review of the basic idea of an electron shuttle (Sec.~\ref{sec:Phenomenology}) followed by mathematical descriptions of the system at the different levels mentioned above (Sec.~\ref{sec:Modelling}): the fully stochastic model in Sec.~\ref{subsec:Stochastic}, the mean-field approach in Sec.~\ref{subsec:MeanField}, and the intermediate, perturbative model in Sec.~\ref{subsec:MSPT}. The dynamics at the different levels are discussed and compared in Sec.~\ref{subsec:Dynamics}. In Sec.~\ref{sec:Thermodynamics} the first and second laws of thermodynamics are derived at the different levels of description (Secs.~\ref{subsec:StochasticThermodynamics}-\ref{subsec:MSPTTD}), followed by a discussion of the thermodynamic behavior of the electron shuttle (Sec.~\ref{subsec:DiscussionTD}). Finally, in Sec.~\ref{sec:Conclusion}, we discuss our findings and point out future applications. 

\section{\label{sec:Phenomenology}Phenomenology}
\begin{figure}
\begin{center}
\includegraphics[width=0.45\columnwidth]{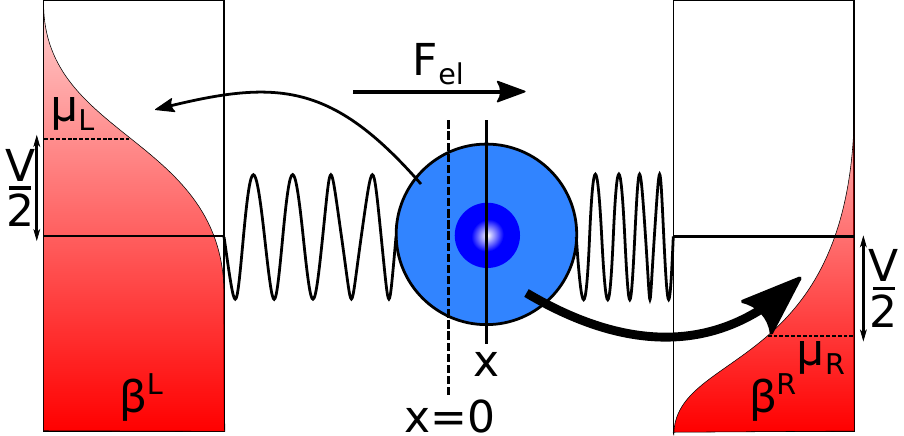}
\includegraphics[width=0.45\columnwidth]{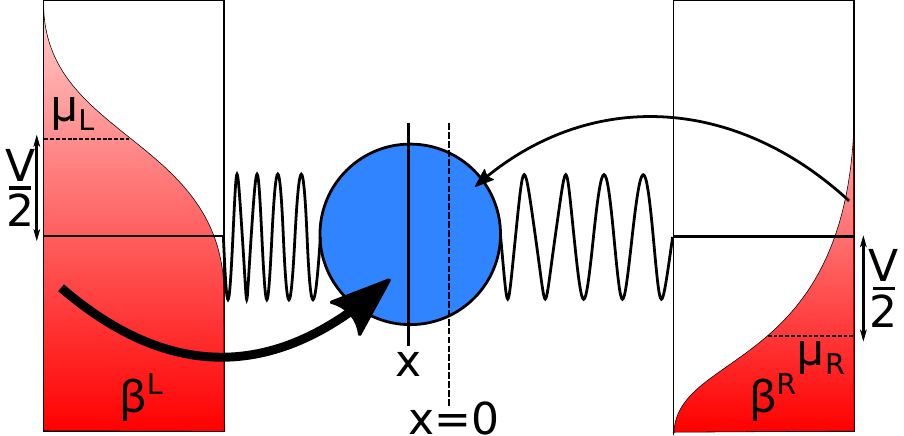}
\end{center}
\caption{Illustration of the model and shuttling mechanism: A single-level dot is coupled to two electronic reservoirs with chemical potentials $\mu^{\text{L}}$ and $\mu^{\text{R}}$ and (inverse) temperatures $\beta^{L/R}$. If an electron tunnels into the dot ($q=1$) the electrostatic force generated by the bias voltage between the leads pushes the oscillator towards the right lead (left figure). If the dot is unoccupied ($q=0$) only the oscillator restoring force acts on the system, pushing the oscillator back towards the center. For weak friction the shuttle may pass the center and approach the left lead, closing the cycle (right figure). The tunnelling rates into and out of the left and right reservoirs depend exponentially on the position $x$ of the shuttle, such that electrons tunnel more likely between the QD and the closer lead (see main text), as indicated by the thickness of the arrows. Additionally, the shuttle is subject to thermal noise (not shown).}
\label{fig:SecModelFig1}
\end{figure}

In this section we explain the basic mechanism of the electron shuttle (see also Fig.~\ref{fig:SecModelFig1}) before introducing the mathematical descriptions in Sec.~\ref{sec:Modelling}. The shuttle is composed of a metallic grain \cite{MoskalenkoEtAlPRB2009, MoskalenkoEtAlNanotechnology2009, KonigWeigAPL2012} or molecular cluster \cite{ScheibleBlickAPL2004, ScorranoCarcaterraMSSSP2013} and a nanomechanical oscillator (e.g. a cantilever \cite{IsacssonPRB2001} or an oscillating molecule \cite{ParkEtAlNature2000, GalperinEtAlJoP2007}), which hosts the grain or cluster and can oscillate. Furthermore, the shuttle is tunnel-coupled to two leads, such that electrons can jump between the leads and the grain. Here, the rate of tunnelling depends on the position of the shuttle -- the closer the shuttle is to the lead, the larger is the rate of tunnelling. A bias voltage applied to the two leads then generates an electric field. The shuttle mechanism works as follows: When the shuttle is close to the reservoir with higher chemical potential, electrons are loaded onto the grain. The electrostatic force due to the electric field between the leads pushes the negatively charged shuttle towards the reservoir with lower chemical potential similar to a charged particle in a capacitor (see Fig.~\ref{fig:SecModelFig1} left). As the shuttle approaches the positively biased reservoir with lower chemical potential, the electrons are unloaded from  the grain, leaving it uncharged. Due to the oscillator restoring force the shuttle returns (see Fig.~\ref{fig:SecModelFig1} right) and the cycle starts again. Above a critical value of the applied bias voltage the damping due to friction is overturned by the electrostatic force. As a result, oscillations of the shuttle are sustained and in each cycle a number of electrons are transported from one lead to the other.

\section{\label{sec:Modelling} Modelling}
In this section we discuss different levels of description of the electron shuttle, i.e., a fully stochastic description in Sec.~\ref{subsec:Stochastic}, a mean-field approximate description in Sec.~\ref{subsec:MeanField} and a perturbative description based on time scale separation in Sec.~\ref{subsec:MSPT}. We will then compare and discuss the dynamics of the system at the different levels in Sec.~\ref{subsec:Dynamics}.

In the literature there exist proposals to describe the electron shuttle fully quantum mechanically \cite{BoeseSchoellerEPL2001, ArmourMacKinnonPRB2002, MccarthyEtAlPRB2003, NovotnyEtAlPRL2003, NovotnyEtAlPRL2004, UtamiEtAlPRB2006}, semiclassically \cite{FedoretsEtAlEPL2002, UtamiEtAlPRB2006, NoceraEtAlPRB2011} or fully classically  \cite{GorelikEtAlPRL1998, IsacssonEtAlPhysicaB1998, WeissZwergerEPL1999, NordEtAlPRB2002}. In this work we describe the system classically, which is justified if the intra-grain electronic relaxation time is much shorter than the tunnelling charge relaxation time \cite{ShekhterEtAlJoP2003}. The latter is the case for an experimental realization of the shuttle with a gold grain \cite{MoskalenkoEtAlPRB2009} and is sometimes referred to as classical shuttling of particles \cite{ShekhterEtAlJoP2003}. The underlying mechanism (tunneling of electrons via Fermi's golden rule), is nevertheless intrinsically quantum.

\subsection{\label{subsec:Stochastic}Fully stochastic description}
We here introduce the specific model of an electron shuttle considered in this work.
In contrast to the original proposal \cite{GorelikEtAlPRL1998} we idealize the quantum dot (QD) by assuming Coulomb blockade.
That is, we assume the QD can accept no more than a single excess electron, due to Coulomb repulsion.
Hence the QD charge state can take the two values $q=0$ (empty) and $q=1$ (occupied).
In this scenario electrons can be transferred one by one between the two reservoirs \cite{GiaeverZellerPRL1968, KulikShekhter1975, 
AverinLikharevJLTP1986}. We use this simplified model of a single electron shuttle for illustrational and numerical purposes, but the phenomenology discussed in Sec.~\ref{sec:Phenomenology} does not change if multiple electrons are allowed on the QD.

The QD with on-site energy $\varepsilon$ is hosted by a nanomechanical oscillator. In the following we will refer to this 
combined system of QD and oscillator as a ``shuttle''. We describe the movement of the oscillator in one dimension with position $x\in\mathbb R$ and velocity $v \in \mathbb R$. The charge $q$ of the shuttle (setting the electron charge $e\equiv 1$) can change due to electron tunnelling with one 
of the two electronic leads, left or right, with chemical potentials $\mu^{\text{L}} = \varepsilon+V/2$ and 
$\mu^{\text{R}} = \varepsilon - V/2$ for the left and right reservoir, respectively. The bias voltage between the two fermionic reservoirs is then given by $V = \mu^{\text{L}} - \mu^{\text{R}}$. 

The QD charge state and the motion of the shuttle are coupled by the electric field that is generated by the bias voltage and assumed to be homogeneous between the leads \cite{GorelikEtAlPRL1998}.
Thus an electrostatic force $F_{\text{el}}=\alpha V q$ acts on the shuttle when it is charged ($q=1$), pushing it towards the reservoir with lower chemical potential (see Fig.~\ref{fig:SecModelFig1} left).
Here $\alpha$ is an effective inverse distance between the leads.
When there is no excess electron on the shuttle ($q=0$) this electrostatic force is absent.

The mechanical vibrations of 
the QD are modelled as a harmonic oscillator with an effective mass $m$ \cite{GorelikEtAlPRL1998, 
ParkEtAlNature2000, GalperinEtAlJoP2007, ShekhterEtAlJoP2003}. From a classical point of view, this restoring force can be explained through interactions between the shuttle, its anchor and the leads, which can be approximated by a harmonic potential 
\cite{ParkEtAlNature2000}. The restoring force acting on the shuttle is then given by $F_{\text{harm}}=-kx$ with spring 
constant $k$, and the shuttle is damped by $F_{\text{damp}}=-\gamma v$ with friction coefficient $\gamma$. We assume underdamped motion to enable the possibility of oscillatory shuttling.
Additionally, we connect the oscillator to its own heat bath at inverse temperature $\beta^{\text{osc}}$ stemming from a dissipative medium in equilibrium. The state of the shuttle is described by the triple $(x,v,q)$. 
Combining the electron jumps with the underdamped oscillations and thermal fluctuations, we describe the dynamics of the shuttle by a generalized FPE
\begin{equation}
\label{eq:FullFPE}
\begin{aligned}
\frac{\partial p}{\partial t}= \left[-v\frac{\partial}{\partial x} +\frac{\partial}{\partial v}\left(\frac{k}{m} x+\frac{\gamma}{m} v -\frac{\alpha V}{m} q\right)+ D \frac{\partial^2}{\partial v^2}\right] p + \sum\limits_{q'\nu}R^\nu_{qq'}(x)p(x,v,q^\prime,t).
\end{aligned}
\end{equation}
Here, $p \equiv p(x,v,q,t)$ denotes the joint probability density to find the shuttle at position $x$ and velocity $v$ with $q\in\{0,1\}$ electrons at time $t$, and we have introduced a velocity diffusion coefficient 
$D=\gamma/(\beta^{\text{osc}}m^2)$.
The first term of Eq.~(\ref{eq:FullFPE}) describes the underdamped evolution of the oscillator in the potential $U_q(x) = \frac{1}{2}kx^2-\alpha V q x$, at fixed electron charge $q$.
The second term couples the mechanical variables ($x,v$) to the charge state ($q$), through a rate equation describing transitions from state $q'$ to state $q$, corresponding to the tunnelling of electrons between the QD and the fermionic lead $\nu\in\{L,R\}$. The transition rates $R_{qq'}^\nu(x)$ are given by 
\begin{equation}
\label{eq:Rate}
\begin{aligned}
R^{\text{L}}_{10}(x)&= \Gamma e^{- x/\lambda}f^{\text{L}}(\varepsilon - \alpha V x ),\\
R^{\text{R}}_{10}(x)&= \Gamma e^{+ x/\lambda}f^{\text{R}}(\varepsilon - \alpha V x ),\\
R^{\text{L}}_{01}(x) &= \Gamma e^{- x/\lambda}\left[1-f^{\text{L}}(\varepsilon - \alpha V x)\right],\\
R^{\text{R}}_{01}(x) &= \Gamma e^{+ x/\lambda}\left[1-f^{\text{R}}(\varepsilon - \alpha V x)\right],
\end{aligned}
\end{equation}
and $R_{qq}^\nu(x) = -\sum_{q'\neq q}R_{q'q}^\nu(x)$, which guarantees the conservation of probability. Here, $\Gamma$ denotes the bare transition rate, which for simplicity we take to be equal for the two fermionic reservoirs. The probability for quantum mechanical tunnelling is exponentially sensitive to the tunnelling distance, such that the tunnelling amplitudes are modulated by the dimensionless displacement $x/\lambda$ of the center of mass of the shuttle \cite{GorelikEtAlPRL1998, LaiEtAlJoP2012, LaiEtAlJoP2013, FedoretsEtAlPRL2004, FedoretsEtAlEPL2002}, where $\lambda$ is a characteristic tunnelling length. Furthermore, the rates depend on the probability of an electron (hole) with a matching energy in the reservoir, i.e., on the Fermi distribution  $f^\nu(\omega) \equiv \left[\exp(\beta^\nu(\omega-\mu^\nu))+1\right]^{-1}$ with inverse temperature $\beta^\nu$. 
Note that the quantity $\varepsilon - \alpha V x$ enters the Fermi functions in Eq.~(\ref{eq:Rate}), as the energy of the shuttle depends on both the QD energy $\varepsilon$ and the electrostatic potential $- \alpha V x$ (see also Sec.~\ref{subsec:StochasticThermodynamics}).

Eq.~(\ref{eq:FullFPE}) describes a system connected to three reservoirs at generally different temperatures: a thermal reservoir of the oscillator at inverse temperature $\beta^{\text{osc}}$ and two fermionic reservoirs with inverse temperature $\beta^\nu$ and chemical potential $\mu^\nu$. While the derivations in this work are general, when solving the dynamics numerically we will focus on the case of equal temperatures, $\beta^{\text{osc}}=\beta^\nu = \beta$. Nonequilibrium conditions then arise solely due to the applied bias voltage, i.e., $\mu^{\text{L}} \neq \mu^{\text{R}}$.

Since the space of dynamical variables defined by the triple $(x,v,q)$ is large, solving Eq.~(\ref{eq:FullFPE}) numerically is expensive. We therefore turn to the trajectory representation of the single electron shuttle.
The coupled stochastic differential equations
\begin{align}
dx &= v dt, \label{eq:StochasticEquations1}\\
m dv &= (-kx -\gamma v + \alpha V q)dt + \sqrt{2 Dm^2}dB(t), \label{eq:StochasticEquations2}\\
dq &= \sum\limits_\nu dq^\nu= \sum\limits_{\nu q'}(q'-q)dN_{q'q}^\nu(x,t).\label{eq:StochasticEquations3}
\end{align}
produce the FPE~(\ref{eq:FullFPE}) at the ensemble level, as we show in \ref{secApp:Equivalence} by explicitly looking at the evolution of averages. In Eq.~(\ref{eq:StochasticEquations2}) the thermal fluctuations are taken into account by a Wiener process $dB(t)$ with zero mean $\mathbb E\left[dB(t)\right] = 0$ and variance $\mathbb E\left[(dB(t))^2\right] = dt$.
Here, $\mathbb E\left[\bullet\right]$ denotes an average of the stochastic process. Eqs.~(\ref{eq:StochasticEquations1}) and (\ref{eq:StochasticEquations2}) represent for a fixed $q$ the Langevin equation of an underdamped particle moving in the shifted harmonic potential $U_q(x)$. 
Eq.~(\ref{eq:StochasticEquations3}) describes changes in the charge state $q$ due to the stochastic tunnelling of electrons.
The independent Poisson increments $dN_{q'q}^\nu(x,t)\in\{0,1\}$ obey the statistics:
\begin{equation}
\label{eq:PoissonIncrement}
\begin{aligned}
\mathbb E\left[dN_{q'q}^\nu(x,t)\right] &= R^\nu_{q'q}(x) dt,\\
dN_{q'q}^\nu(x,t) dN_{\tilde qq}^{\tilde \nu}(x,t) &= \delta_{q'\tilde q}\delta_{\nu\tilde \nu} dN_{q'q}^{\nu}(x,t).
\end{aligned}
\end{equation}
The first equation specifies that the average number of jumps into state $q'$ from a state $q$ in a time interval $dt$ is given by the tunnelling rate $R^\nu_{q'q}(x)$.
The second line in Eq.~(\ref{eq:PoissonIncrement}) enforces that only one tunnelling event per time interval can occur, i.e., either all $dN_{q'q}^\nu(x,t)$ are zero or $dN_{q'q}^\nu(x,t)=1$ for precisely one set of indices $q$, $q'$ and $\nu$. 

Well-known models emerge as a simple limit of our description. First, for $\alpha \to 0$ the motion of the oscillator becomes independent of the charge state $q$, and Eqs.~(\ref{eq:StochasticEquations1}) and (\ref{eq:StochasticEquations2}) describe a simple underdamped harmonic oscillator. However, the tunnelling of electrons still depends on $x$ [see Eq.~(\ref{eq:Rate})] and therefore the QD remains coupled to the oscillator. Second, a complete decoupling of the QD and the oscillator is achieved in the limit $\lambda\to\infty$ \emph{and} $\alpha \to 0$.
In that case the QD  coupled to the fermionic leads describes the well known single electron transistor (SET) \cite{BonetEtAlPRB2002, BagretsEtAlPRB2003, HarbolaPRB2006}.
 
\subsection{\label{subsec:MeanField}Mean-field approximation}
In order to understand the nonlinear dynamics of the compound system of QD and oscillator, we first look at the mean-field equations derived from the full stochastic evolution.
From the FPE, Eq.~(\ref{eq:FullFPE}), we obtain for the ensemble averaged position $\left<x\right>$ and velocity $\left<v\right>$:
\begin{equation}
\label{eq:BeforeMeanField}
\begin{aligned}
\frac{d}{dt}\left< x\right> &= \left< v\right>,\\
m\frac{d}{dt}\left<v\right> &= -k \left<x\right>-\gamma  \left<v\right> +\alpha V p_1.
\end{aligned}
\end{equation}
Here and throughout the paper,
\begin{equation}
\left<\bullet\right> = \int dx dv\sum_q \bullet \, p(x,v,q,t)
\end{equation}
denotes an ensemble average, and
\begin{align}
p_0 &= \int dx dv \, p(x,v,0,t) \\
p_1 &= \int dx dv \, p(x,v,1,t)
\end{align}
are the probabilities for the QD to be empty and occupied, respectively.
Eqs.~(\ref{eq:BeforeMeanField}) are exact, but in order for them to form a closed set we need an expression for $dp_1/dt$.
Integrating Eq.~(\ref{eq:FullFPE}) over $x$ and $v$ we obtain
\begin{equation}
\frac{\partial}{\partial t}p_{q} = \int dx \, dv \sum\limits_{q'\nu}R^\nu_{qq'}(x)p(x,v,q^\prime,t) = \sum\limits_\nu \left<R^\nu_{qq'}(x)\right>. 
\end{equation}
Due to the nonlinearity of the tunnelling rates with respect to $x$ [see Eq.~(\ref{eq:Rate})] we approximate 
\begin{equation}
\label{eq:MFApproximation}
\left<R^\nu_{qq'}(x)\right> \approx R^\nu_{qq'}\left(\left<x\right>\right),
\end{equation} 
and we refer to this as the mean-field (MF) approximation.
Note that if the tunnelling rates $R^\nu_{qq'}(x)$ were linear in $x$, Eq.~(\ref{eq:MFApproximation}) would be an equality and Eq.~(\ref{eq:BeforeMeanField}) would be closed without the MF approximation. 

The MF approximation is thus described by the nonlinear differential equations
\begin{align}
\frac{d}{dt}\bar x &= \bar v,\label{eq:MeanField1}\\
m\frac{d}{dt}\bar v &= -k\bar x-\gamma \bar v +\alpha V \bar q, \label{eq:MeanField2}\\
\frac{d}{dt} \bar \pb &= \sum\limits_\nu R^\nu(\bar x)\bar \pb ,\label{eq:MeanField3}
\end{align}
where $\bar \pb \equiv \left(\bar p_0,\bar p_{1}\right)^\intercal$ and $\bar q \equiv \bar p_{1}$. The entries of the rate matrix $R^\nu(\bar x)$ 
are definded by Eq.~(\ref{eq:Rate}), i.e., $\left[R^\nu(\bar x)\right]_{qq'} \equiv R^\nu_{qq'}(\bar x)$. The overbars 
denote that the quantities are governed by MF equations. Within the MF description, the QD is still described by a probability and therefore behaves stochastically, whereas the oscillator is fully deterministic.  

The temperature of the oscillator bath, $\beta^{\text{osc}}$, does not appear in Eqs.~(\ref{eq:MeanField1})-(\ref{eq:MeanField3}).
In effect the MF approximation describes a macroscopic system for which thermal fluctuations are negligible, as would be expected in the limit of large oscillator mass.
In this limit the oscillation period $2\pi\sqrt{m/k}$ becomes much longer than the time scale associated with changes in the charge state $q$, hence the charge state can be replaced by its local-in-time average, as reflected in Eq.~(\ref{eq:MeanField2}).
Eqs.~(\ref{eq:MeanField1})-(\ref{eq:MeanField3}) are identical to those found in the original proposal of Gorelik \textit{et al.} \cite{GorelikEtAlPRL1998} for the case of one excess electron.

\subsection{\label{subsec:MSPT}Multiple scale perturbation theory}
To improve on the MF approximation, which only captures the average dynamics of the electron shuttle, we can perturbatively solve the FPE, Eq.~(\ref{eq:FullFPE}),
by assuming a  separation of time scales between the short dwell-time of electrons and the slow movement of the oscillator, and applying multiple scale (MS) perturbation theory \cite{KevorkianColeBook1996, NayfehBook2008, BenderOrszagBook2013}.  Specifically, we assume that during one oscillation there are many electron tunnelling events: $\Gamma \gg \sqrt{k/m}$.
We provide details of the MS calculation in \ref{secApp:MultipleScale}, and summarize the result here.

Working to first order in the perturbation we obtain
\begin{equation}
\label{eq:ApproxedProbability}
p(x,v,q,t) \approx \pi_q(x)\tilde p(x,v,t),
\end{equation}
where the vector $(\pi_0(x),\pi_1(x))^T$ denotes the stationary state of the rate matrix $R(x)=\sum_\nu R^\nu(x)$, i.e.\ it is the right eigenvector corresponding to the zero eigenvalue, normalized to unity: $\sum_q \pi_q(x) =1$. The probability density to find the oscillator at position $x$ with velocity $v$ at time $t$ is given by $\tilde p(x,v,t)$, which obeys the FPE:
\begin{equation}
\label{eq:FokkerPlanckHO}
\begin{aligned}
\frac{\partial \tilde p }{\partial t}  = \left[-v \frac{\partial}{\partial x} + \frac{\partial}{\partial v}\left(\frac{k}{m}x + \frac{\tilde \gamma(x)}{m}v - \frac{\alpha V }{m}q_{\text{eq}}(x) \right)\right]\tilde p + \tilde D(x) \frac{\partial^2 \tilde p}{\partial v^2}.
\end{aligned}
\end{equation}
Here, $q_{\text{eq}} (x) \equiv \pi_1(x)$ is the instantaneous stationary charge of the QD given by
\begin{equation}
q_{\text{eq}}(x) = \frac{f^{\text{L}}(\varepsilon-\alpha V x) - f^{\text{R}}(\varepsilon - \alpha V x)}{1+e^{2x/\lambda}}+f^{\text{R}}(\varepsilon-\alpha V x).
\end{equation}
As seen in Eq.~(\ref{eq:FokkerPlanckHO}), the effects of the electronic degrees of freedom on the evolution of the oscillator are incorporated into an effective potential $U_{\text{eff}}(x)$, along with position dependent friction and diffusion coefficients:
\begin{align}
U_{\text{eff}}(x) &= kx-\alpha Vq_{\text{eq}}(x), \\
\label{eq:DefinitionGammaBar}
\tilde \gamma(x) &= \gamma - \frac{\alpha V}{ \chi(x)}\frac{\partial q_{\text{eq}}(x)}{\partial x}, \\
\label{eq:DefinitionDBar}
\tilde D(x) &=  D-\frac{\alpha^2 V^2 q_{\text{eq}}(x)}{m^2 \chi(x) }\left[1-q_{\text{eq}}(x)\right].
\end{align}
where $\chi(x)=-2\Gamma \text{cosh}(x/\lambda)$ is the non-zero eigenvalue of $R(x)$.

The zeroth order perturbation (see \ref{secApp:MultipleScale}) corresponds to an adiabatic approximation, i.e., infinite time scale separation, $\Gamma\rightarrow\infty$.
In that limit we have $\tilde\gamma(x)\rightarrow\gamma$ and $\tilde D(x)\rightarrow D$, and Eq.~(\ref{eq:FokkerPlanckHO}) describes underdamped Brownian motion in an effective potential $U_{\text{eff}}(x)$, at inverse temperature $\gamma/(Dm^2) = \beta^{\text{osc}}$.
In this situation, detailed balance is satisfied and the oscillator relaxes to an effective equilibrium state, with no self-sustained oscillations.
By contrast, in the first order perturbation represented by Eq.~(\ref{eq:FokkerPlanckHO}), the $x$-dependence of $\tilde\gamma/(\tilde D m^2)$ breaks detailed balance, giving rise to non-equilibrium behavior and allowing for the possibility of self-oscillations.

To solve Eq.~(\ref{eq:FokkerPlanckHO}) approximately, we parametrize $x$ and $v$ by the energy $\mathcal E$ and the oscillation phase $\theta$, 
\begin{equation}
\label{eq:Transformation}
x = \sqrt{\frac{2 \mathcal E}{k}}\sin \theta,~v=\sqrt{\frac{2 \mathcal E}{m}} \cos \theta, 
\end{equation}
such that $\frac{1}{2}kx^2+\frac{1}{2}mv^2 = \mathcal E$, and we assume that the probability density does not depend on the phase \cite{BlanterEtAlPRL2004}: $\hat p(\mathcal E,\theta,t) \approx \hat p(\mathcal E,t)$. With the transformations of Eq.~(\ref{eq:Transformation}) and the latter assumption we find a FPE for the energy distribution by averaging over the angle $\theta$ (see \ref{secApp:PolarCoordinates}):
\begin{equation}
\label{eq:FPEEnergy}
\frac{\partial}{\partial t}\hat p(\mathcal E,t) = \frac{\partial}{\partial \mathcal E}
\left[2 \mathcal E \left(\frac{\hat \gamma }{m} + m \hat D\frac{\partial }{\partial \mathcal E}\right) \hat p(\mathcal E,t) \right],
\end{equation}
where $\hat p(E,t)$ is the transformed probability distribution $\tilde p(x,v,t)$. The effective friction and diffusion parameters take after the transformation the form
\begin{equation}
\label{eq:TransformedGammaE}
\begin{aligned}
\hat \gamma (\mathcal E) &= \frac{1}{2\pi}\int\limits_0^{2\pi} d\theta~\tilde \gamma (x) \cos^2 \theta, \\
\hat D (\mathcal E) &= \frac{1}{2\pi}\int\limits_0^{2\pi} d\theta~\tilde D (x) \cos^2 \theta.
\end{aligned}
\end{equation}
Solving for the steady state of Eq.~(\ref{eq:FPEEnergy}), i.e.\ $\partial \hat p/\partial t = 0$, we get \cite{RiskenBook1996}
\begin{equation}
\label{eq:SolutionMSPT}
\hat p_{ss}(\mathcal E) = \mathcal N \exp\left(-\int\limits_0^\mathcal E \frac{\hat \gamma (\mathcal E')}{m^2 \hat D(\mathcal E')}d\mathcal E'\right),
\end{equation}
where $\mathcal N$ is a normalization constant.

\subsection{\label{subsec:Dynamics}Dynamics on the different levels of description}
In this section we discuss and compare the dynamical behaviour of the single electron shuttle on the different levels of description introduced before. All numerical results in this section are obtained using the parameter values specified in \ref{secApp:ComputationalMethods}.

\subsubsection{Stochastic dynamics} 
\begin{figure}
\begin{center}
\includegraphics[width=0.7\columnwidth]{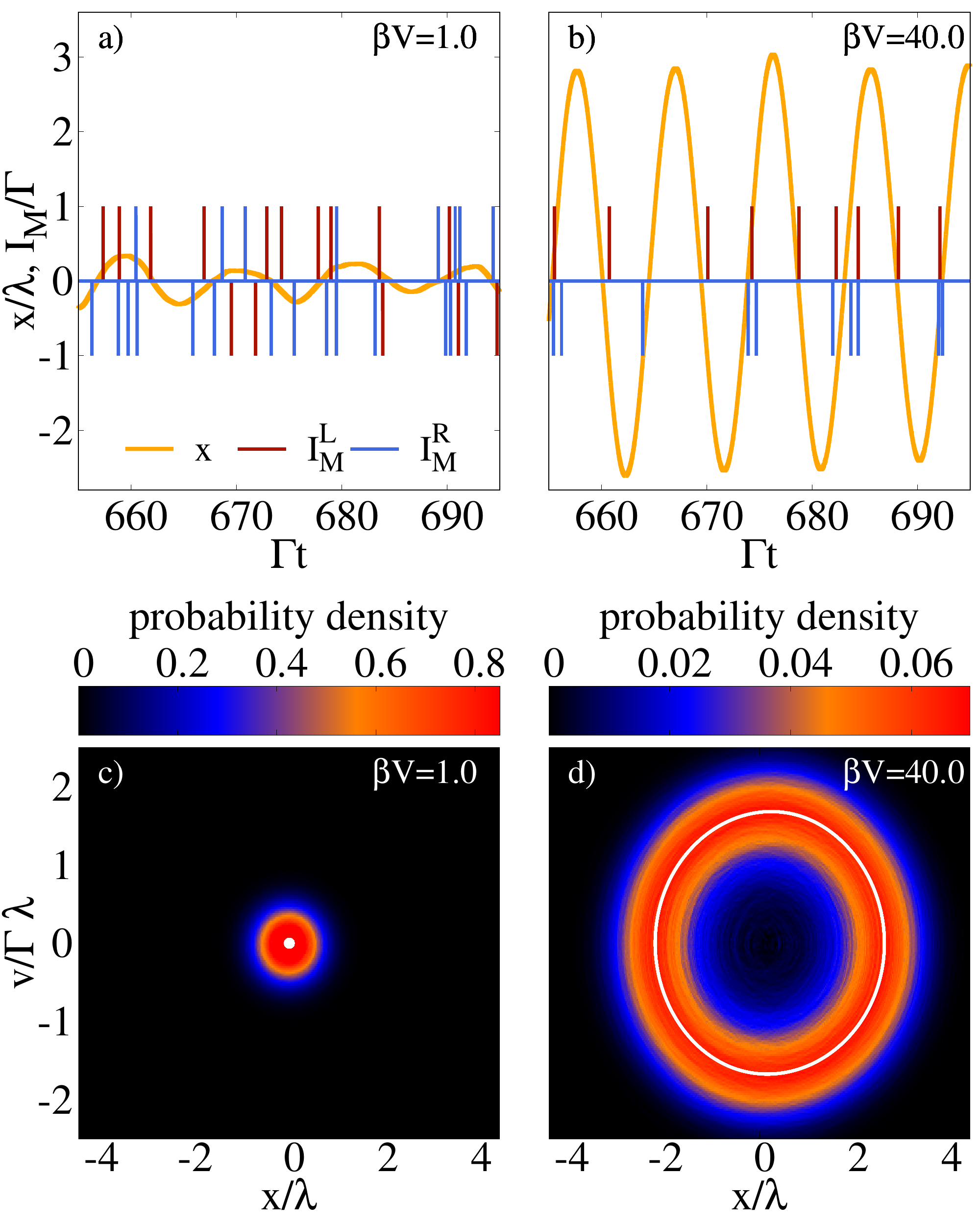}
\end{center}
\caption{Top: Exemplary trajectories of the position (orange) together with electron jumps between the QD and the left (red) and right (blue) reservoir for a) $\beta V = 1.0$ and b) $\beta V = 40.0$ showing clearly the shuttling for a large bias voltage. Bottom: Probability density of the oscillator in phase-space for c) $\beta V = 1.0$ and d) $\beta V = 40.0$ simulated from Eqs.~(\ref{eq:StochasticEquations1})-(\ref{eq:StochasticEquations3}) (see also \ref{secApp:ComputationalMethods}). The circular orbit indicates self-oscillations. The white dot and circle correspond to MF solutions [see Eqs.~(\ref{eq:MeanField1})-(\ref{eq:MeanField3})].}
\label{fig:DensityPlots}
\end{figure}

We start by looking at the fully stochastic model, given by Eq.~(\ref{eq:FullFPE}).
Rather than solving the FPE directly, we generated stochastic trajectories evolving under the Langevin Eqs.~(\ref{eq:StochasticEquations1}) - (\ref{eq:StochasticEquations3}).
Fig.~\ref{fig:DensityPlots} shows trajectory segments $x(t)$ for two 
different values of the applied bias voltage: $\beta V=1.0$ [Fig.~\ref{fig:DensityPlots} a)] and $\beta V = 40.0$ [Fig.~\ref{fig:DensityPlots} b)]. The two 
figures show quite different behaviour of the stochastic position (orange) as well as the tunnelling of electrons 
schematically indicated by red (left lead) and blue (right lead) bars. Here, a negative value of $I_{\text{M}}^\nu = dq^\nu/dt$ denotes the 
jump of an electron from the QD into the reservoir $\nu$ ($dq^\nu = -1$) whereas a positive value indicates the reverse 
process ($dq^\nu =1$). Note that the stochastic current $I_{\text{M}}^\nu$ along a trajectory shows up as delta-peaks. In Fig.~\ref{fig:DensityPlots} a) and b) we plot $dq^\nu$ for clearness of the figures.
For a small bias voltage [panel a)] tunnelling events are frequent and the position $x(t)$
oscillates irregularly around the equilibrium position. In contrast, Fig.~\ref{fig:DensityPlots} b) shows regular oscillations 
of the position and fewer tunnelling events. Also, the tunnelling events in Fig.~\ref{fig:DensityPlots} b) are synchronized with the shuttling:
during each period of oscillation, the empty shuttle picks up one electron from the left reservoir (red bar) as it moves past the origin in a leftward direction, $dx/dt<0$,
and it releases that electron to the right reservoir on its way back (blue bar), as it moves past the origin in a rightward direction, $dx/dt>0$.
(Typically, immediately after releasing the electron the shuttle picks up another electron from the left reservoir and quickly delivers it to the right reservoir.)
This behaviour reflects the mechanism of single electron shuttling discussed in Sec.~\ref{sec:Phenomenology}. 

The directed shuttling of electrons coincides with self-oscillation, as illustrated in Fig.~\ref{fig:DensityPlots} c) and d), which show
the stationary probability density of the oscillator, $p(x,v) = \sum_q p(x,v,q)$, obtained by 
simulating a long trajectory evolving under Eqs.~(\ref{eq:StochasticEquations1})-(\ref{eq:StochasticEquations3}) and assuming ergodicity (see \ref{secApp:ComputationalMethods}).
For small bias voltage [panel c)], the probability density is peaked close to the origin, showing no sign of regular oscillations.
When the applied voltage is larger [panel d)], the probability density is concentrated around a circular orbit, revealing self-oscillatory harmonic motion with some amplitude and phase noise. These behaviours are consistent with the trajectories shown in Fig.~\ref{fig:DensityPlots} a) and b), respectively, and they suggest that there exists a value of the applied bias voltage $V$ above which the shuttle oscillates, as we will discuss below.
Similar oscillator distributions in phase space have been observed for Wigner functions in semiclassical descriptions of the single electron shuttle \cite{NovotnyEtAlPRL2003, NovotnyEtAlPRL2004}. 
Note that $V$ enters the equations governing the dynamics via the coupling to the electrostatic field \emph{and} via the chemical potentials [see Eq.~(\ref{eq:Rate})]. 

\begin{figure}[h]
\begin{center}
\includegraphics[width=0.7\columnwidth]{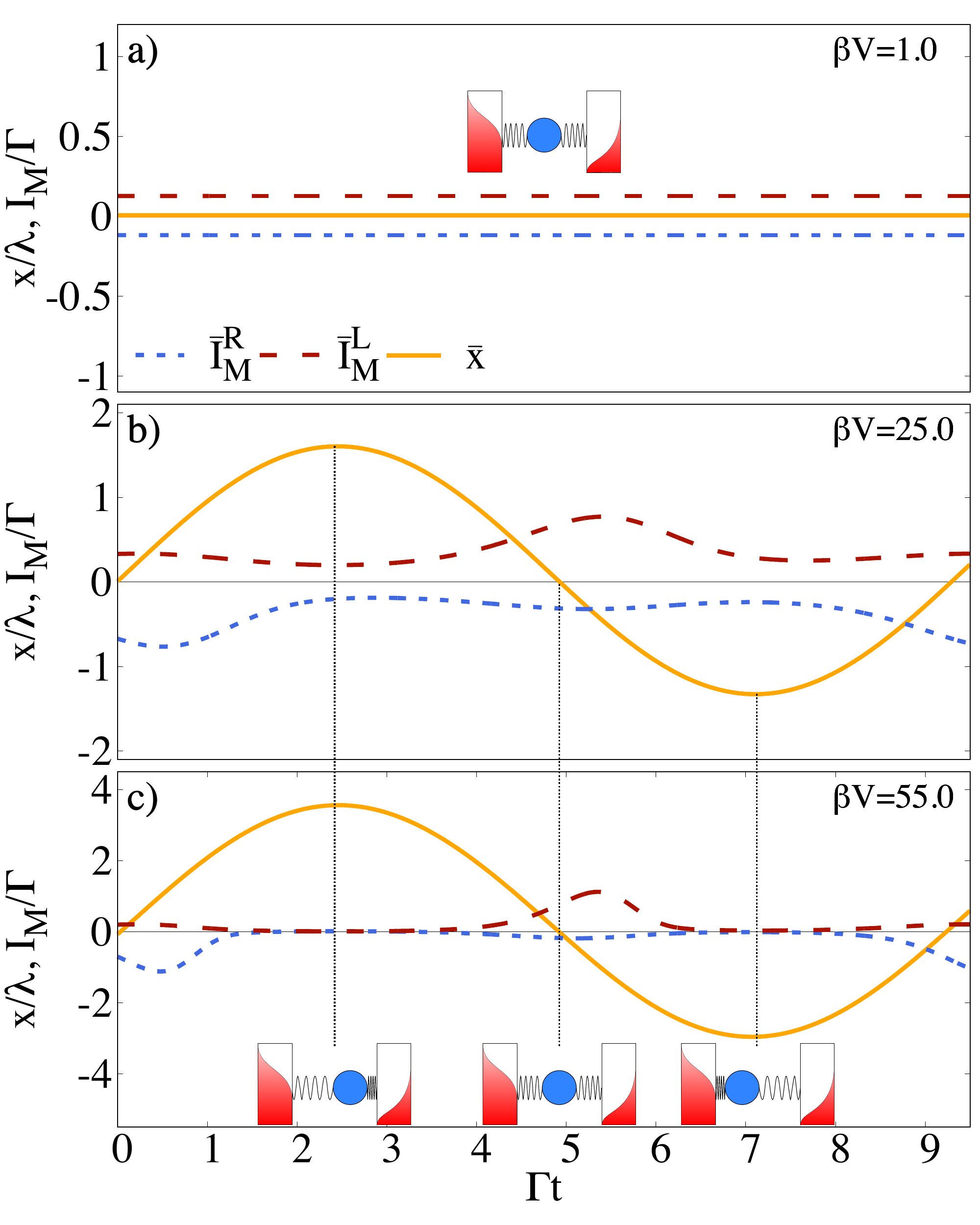}
\end{center}
\caption{MF position $\bar x$ (orange solid) as well as right (blue dotted) and left (red dashed) matter current, $\bar I_{\text{M}}^{\text{R}}$ and $\bar I_{\text{M}}^{\text{L}}$, during one period in the asymptotic limit. From a) to c) the bias voltage $V$ is increased. Below $\beta\bar V_{\text{cr}}=15.0$ the system is equivalent to a SET and $\bar x = \alpha V \bar q/k$ is constant as indicated also by the illustration in a) [regime (I)]. Above the critical voltage the system oscillates and after a crossover regime (II) [panel b)] the system acts as an electron shuttle (III) transporting one electron per cycle. The illustrations in c) indicate the position of the oscillator during the cycle.}
\label{fig:Dynamics}
\end{figure}

\subsubsection{Mean-field dynamics} 
\label{subsec:mfd}
We now turn to the mean-field (MF) dynamics. The white dot and circle in Fig.~\ref{fig:DensityPlots} c) and d) correspond to the solutions of the MF model given by 
Eqs.~(\ref{eq:MeanField1})-(\ref{eq:MeanField3}). As we can see the MF solutions coincide very well with the stochastic phase-space 
distribution.
As shown in previous extensive studies \cite{UtamiEtAlPRB2006}, when the parameter $V$ crosses a critical value $\bar V_{\text{cr}}$, the MF system undergoes a Hopf bifurcation from a stable fixed point to a stable limit cycle.
For our choice of parameters the bifurcation takes place at $\beta\bar V_{\text{cr}}=15.0$ (see  \ref{subsec:Hopf bifurcation}). Three dynamical regimes can be characterized, as we discuss below.

\textit{Single electron transistor (SET) regime (I):}
The point $(\bar x_{\text{fix}},\bar v_{\text{fix}}) = (\alpha V \bar q /k,0)$ is a fixed point of Eqs.~(\ref{eq:MeanField1})-(\ref{eq:MeanField2}), and 
below the critical value of the applied voltage this fixed point is stable: from any initial conditions the oscillator spirals into this point, hence at steady state the MF system does not oscillate 
[see Fig.~\ref{fig:DensityPlots} c)]. 
In Fig.~\ref{fig:Dynamics} a) we find the steady state solution for the MF position at $\bar x =\bar x_{\text{fix}} = 0.006\lambda$ and electron currents $I_{\text{M}}^{\text{L}} = - \bar I_{\text{M}}^{\text{R}} = 0.122\Gamma$, describing a fixed oscillator and a constant matter current from left to right lead. The electrostatic force cannot overcome friction, and the transition rates $R_{qq'}^\nu(\bar x)$ [see Eq.~(\ref{eq:Rate})] are constant since $\bar x$ is constant at steady 
state. The dynamics of the QD are then described by a 
simple rate equation equivalent to the classical master equation of the SET \cite{SchallerBook2014}, leading to a net electron current $\bar I_{\text{M}} = \bar I_{\text{M}}^{\text{L}} = -\bar I_{\text{M}}^{\text{R}} = \Gamma/2\left[f^{\text{L}}(\varepsilon-\alpha V \bar x_{\text{fix}})-f^{\text{R}}(\varepsilon-\alpha V \bar x_{\text{fix}})\right]\text{sech}\left(\bar x_{\text{fix}}/\lambda\right)$.
Note that at the stochastic level [see Fig.~\ref{fig:DensityPlots} a)] the oscillator is 
not fixed -- only the average position and velocity are equal to the fixed point values.

\textit{Shuttling regime (III):} For a bias voltage $V\gg \bar V_{\text{cr}}$ the system is self-oscillating and therefore acts as a 
shuttle transporting one electron from one lead to the other during each cycle 
[see Fig.~\ref{fig:DensityPlots} b) and Fig.~\ref{fig:Dynamics} c)].
When the shuttle is occupied by an electron, the 
electrostatic force is sufficient to overcome friction, leading to self-sustained oscillations. 
Note that perfect shuttling, i.e., transport of one electron per oscillation, only occurs at very large bias voltages.  For a bias of $\beta \bar V=55.0$ there are still more tunnelling events than from shuttling electrons one by one, which in Fig.~\ref{fig:Dynamics} c) can be seen from the fact that both currents $\bar I_{\text{M}}^\nu$ are finite when $x\approx 0$.  We also see this in the stochastic case very clearly 
[see Fig.~\ref{fig:DensityPlots} b)].

\textit{Crossover regime (II):} 
When $V\approx \bar V_{\text{cr}}$ the system exhibits both SET and shuttle behaviour.
Above the critical bias voltage $\bar V_{\text{cr}}$ the MF fixed point 
is unstable and a small perturbation to the system causes variations in the charge of the QD. The electrostatic force 
acting on these charge variations provides positive feedback on the oscillator and compensates for losses due to 
friction. The asymptotic MF state is characterized by periodic oscillations of the position $\bar x$ and velocity $\bar v$ -- as in the shuttling regime -- as well as charge $\bar q$ and matter currents 
$\bar I_{\text{M}}^{\nu}$ [see Fig.~\ref{fig:Dynamics} b)].
However, throughout the entire period of oscillation the QD is able to exchange electrons with both leads, as in the SET regime.
As the bias voltage is increased, the amplitude of oscillations increases, and the time during which the shuttle exchanges electrons with the reservoirs decreases, and finally only one electron is transferred per cycle, which corresponds to the pure shuttling regime.

\subsubsection{Perturbative dynamics}
To gain further insight into the transition to self-oscillation, we solve the full 
FPE, Eq.~(\ref{eq:FullFPE}), perturbatively by imposing a time scale separation between the rapid tunnelling events of electrons and the 
slow movement of the oscillator (see Sec.~\ref{subsec:MSPT}).  In Fig.~\ref{fig:MSPT} a) we plot the steady state 
probability density $\tilde p_{ss}(\mathcal E)$ obtained from this calculation [see Eq.~(\ref{eq:SolutionMSPT})] for different applied bias voltages (dotted).
We also plot the corresponding energy distributions $p_{ss}(\mathcal E)=\int  \delta(\mathcal E- kx^2/2- mv^2/2)p_{ss}(x,v)dx dv$ determined from numerical simulations of the full stochastic evolution (solid).
The two sets of distributions show similar behaviour: for small voltages the maximum occurs at $\mathcal E=0$ but for larger values of $V$ the distributions are peaked at non-zero values of the energy, corresponding to self-oscillation as discussed earlier. For larger values of the voltage the maximum of the probability density occurs at smaller values of the energy for the stochastic case, when compared with the MS results. This deviation can be understood in terms of the underlying assumption of time scale separation for the MS perturbation theory: As the bias voltage is increased the system transitions from the SET regime (with clear time scale separation) to the shuttling regime (where time scales are comparable).

\subsubsection{Comparison}
Finally, we compare all three levels of description in terms of an order parameter $A$ that quantifies the magnitude of self-oscillation.
In the MF case the position at long times performs oscillations of the form $\bar x(t) = \bar A \cos (\bar \omega t +\bar \varphi_0)+\bar x_{\text{fix}}$, and we choose $A_{\text{MF}}=\bar A$ as our order parameter.
In the stochastic case we consider the probability density at $v=0$, i.e., $p(x,v=0)$, which in the case of large self-oscillations resembles a pair of well-separated peaks [see Fig.~\ref{fig:DensityPlots} d)].
We fit $p(x, v=0)$ to a normalized sum of Gaussians, $g(x)= N\left\{\exp[-(x-c-x_0)^2/2\sigma^2]+\exp[-(x-c+x_0)^2/2\sigma^2]\right\}$ with fit parameters\footnote{We have to include a shift $c$ because the the orbit is not exactly centred around the origin.  This shift is the counterpart of $x_{\text{fix}}$ discussed in the MF context in Sec.~\ref{subsec:mfd}.} $\sigma^2$, $c$, and $x_0$.
We then define the self-oscillation amplitude $A_{\text{FPE}}$ in terms of the value(s) $x$ at which the function $f(x)=g(x+c)$ has a maximum:
when $x_0\le\sigma$, $f(x)$ has a unique maximum at $x=A_{\text{FPE}}=0$, and
when $x_0>\sigma$, $f(x)$ has distinct maxima at $x=\pm A_{\text{FPE}}$.
Note that this definition is not sensitive to small oscillations of the stochastic system, as it gives $A_{\text{FPE}}=0$ when $x_0\le\sigma$, even though the shuttle may be self-oscillating.
Finally, for the MS perturbative solution, we define the self-oscillation amplitude as $A_{\text{MS}} = \sqrt{2 \mathcal E_{max}/k}$, where $\mathcal E_{max}$ is the value of $\mathcal E$ at which the function $\hat p_{ss}(\mathcal E)$ is maximized [see Eqs.~(\ref{eq:Transformation}) and (\ref{eq:SolutionMSPT})].
Similarly to $A_{\text{FPE}}$, and for the same reason, $A_{\text{MS}}$ is not sensitive to small oscillations.

\begin{figure}[h]
\begin{center}
\includegraphics[width=0.7\columnwidth]{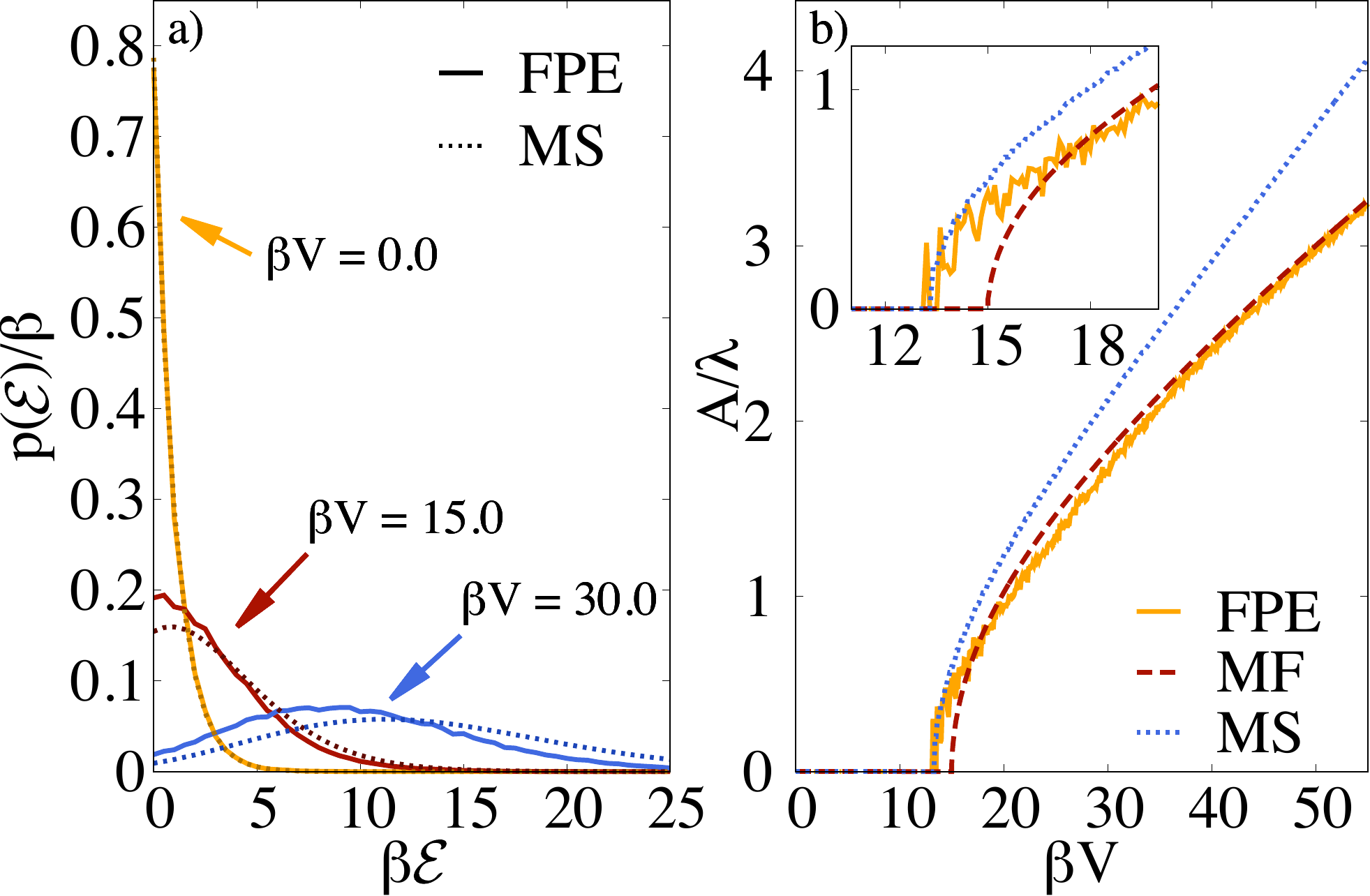}
\end{center}
\caption{Panel a): Numerical solutions of $\hat p_{ss}(E)$ (MS, dotted) together with the equivalent plots of the 
stochastic solution $p(E)$ (FPE, solid) as a function of the energy $E$ of the oscillator. A maximum larger than zero indicates that the 
system is oscillating. For large values of $V$ the MS solution deviates from the stochastic solution due to the break down of the 
time scale separation. Panel b): 'Order parameter' $A$ of the oscillations as a function of $V$ for the FPE (orange solid), the MF (red dashed) and the MS description (blue dotted): All three descriptions predict an onset of oscillation and agree quite well for the chosen set of parameters. The inset shows a zoom into the onset region.}
\label{fig:MSPT}
\end{figure}

In Fig.~\ref{fig:MSPT} b) we show the amplitudes of oscillation $A$ for the different levels of 
description as a function of the applied voltage: FPE (orange solid), MF (red dashed) and MS (blue dotted). All three descriptions show an onset of oscillation at a critical value of the voltage. The specific values are given by $\beta \bar V_{\text{cr}} = 15.0$, $\beta V_{\text{FPE}}^\ast = 13.2$ and $\beta V_{\text{MS}}^\ast = 13.6$. These values are surprisingly close to each other, in particular the MS analysis accurately reflects the onset seen in the full FPE simulations.
Recall, however, that since $A_{\text{FPE}}$ and $A_{\text{MS}}$ are not sensitive to small oscillations, the onset to self-oscillation in the FPE and MS cases may not be as abrupt as suggested by the data in Fig.~\ref{fig:MSPT} b).

As $V$ is increased, the perturbative solution deviates from the stochastic amplitude of oscillation due to the lack of time scale separation, as discussed earlier (see Sec.~\ref{subsec:Dynamics} 3).  On the other hand, for a large bias voltage the MF and stochastic description coincide quite well, as the deterministic component of the dynamics becomes dominant and the fluctuations become less important.
Note that the onset of oscillations in the stochastic case may vary somewhat according to the choice of fitting function $g(x)$.
Also, due to fitting of the probability density, $A_{\text{FPE}}$ is quite noisy close to the onset; see inset of Fig.~\ref{fig:MSPT} b).
Despite these caveats, we see that a transition towards self-oscillation can be identified at all three levels of description.
Next, we investigate whether this transition is reflected in thermodynamic quantities such as chemical work rate, heat flow, and entropy production rate.

\section{\label{sec:Thermodynamics}Thermodynamics}
In this section we formulate the first and second law of thermodynamics at the different levels of description 
introduced above --  stochastic, mean field and multiscale perturbative.
In each case we introduce precise definitions of essential thermodynamic quantities, namely heat, work, and entropy production. With these definitions, we compare the thermodynamic behavior of the electron shuttle at the different levels of description, focusing on the thermodynamic signatures of the onset of spontaneous self-oscillation.

\subsection{\label{subsec:StochasticThermodynamics}Stochastic thermodynamics}
We start by looking at the full stochastic model. The total energy of the coupled system is given by
\begin{equation}
E = \frac{mv^2}{2} + \frac{kx^2}{2} + \varepsilon q-\alpha V x q,
\end{equation}
where the first two terms correspond to the kinetic and potential energy of the harmonic oscillator and the third term is the energy of the QD. The last term describes the interaction energy of the oscillator with the QD, which is given by an electrostatic energy analogous to that of a charged particle in a capacitor with a constant electrostatic field of strength $\alpha V$.

By the first law of thermodynamics, a change in the total energy of the system is due either to exchange of heat or to work performed by (on) the system. The change of total energy in the electron shuttle is expressed as \cite{SekimotoBook2010}
\begin{equation}
\label{eq:firstLaw}
\begin{aligned}
dE &= kx\circ dx + mv\circ dv -\alpha V q \circ dx +(\varepsilon-\alpha V x)\circ dq,
\end{aligned}
\end{equation}
where $\circ$ denotes Stratonovich-type calculus. In the last term, $dq = \sum_\nu dq^\nu$ and $dq^\nu=\sum_{q'}(q'-q)dN^\nu_{q'q}(x,t)$ denotes an electron jump with respect to reservoir $\nu$. The second term involving the velocity can be re-expressed by multiplying Eq.~(\ref{eq:StochasticEquations2}) with $v$. This yields
\begin{equation}
\label{eq:AfterFirstLaw}
\begin{aligned}
mv\circ dv &= (-kxv -\gamma v^2 + \alpha V q v)dt + \sqrt{2\gamma / \beta^{\text{osc}}}v\circ dB(t) \\
&=-kx\circ dx + \alpha V q \circ dx -\gamma v^2 dt \\
&\quad+ \sqrt{2\gamma /\beta^{\text{osc}}}v\circ dB(t).
\end{aligned}
\end{equation}
Inserting Eq.~(\ref{eq:AfterFirstLaw}) into Eq.~(\ref{eq:firstLaw}), we get
\begin{equation}
\label{eq:EnergyTrajectory}
\begin{aligned}
dE &= (\varepsilon - \alpha V x)\circ dq - \gamma v^2 dt + \sqrt{2\gamma /\beta^{\text{osc}} }v\circ dB(t) \\
&= \delta Q^{\text{L}} + \delta Q^{\text{R}} + \delta W^{\text{chem}}+ \delta Q^{\text{osc}}.
\end{aligned}
\end{equation}
Here, we have introduced the chemical work $\delta W^{\text{chem}}=\sum_\nu \mu^\nu dq^\nu$ and the heat flow to the oscillator from its thermal reservoir due to friction and thermal noise $\delta Q^{\text{osc}}=-\gamma v^2dt+\sqrt{2\gamma/\beta^{\text{osc}}}v\circ dB(t)$ \cite{SekimotoBook2010}. 
The remaining terms in Eq.~(\ref{eq:EnergyTrajectory}) are identified as heat exchanged with the reservoir $\nu$, defined as $\delta Q^\nu = (\varepsilon-\alpha V x-\mu^\nu)\circ dq^\nu$.  We use the convention that work performed on the system is positive as is heat transferred from a reservoir into the system. With these definitions of heat and work we can derive a consistent second law as we will show later in this section.

The average change in total energy is given by averaging Eq.~(\ref{eq:EnergyTrajectory}) over many realizations, equivalently by averaging with respect to the probability density (see Sec.~\ref{subsec:Stochastic} and \ref{secApp:Equivalence}):
\begin{equation}
\label{eq:AverageFirstLaw}
\left<\frac{dE}{dt}\right> =\left<\dot Q^{\text{L}}\right> + \left<\dot Q^{\text{R}}\right> + \left<\dot W^{\text{chem}}\right> + \left<\dot Q^{\text{osc}}\right>.
\end{equation}
We note that derivatives with respect to time ($\frac{d}{dt}$) denote exact (or complete) differentials whereas a dot ($\cdot$) denotes inexact ones. 
The average heat absorbed from reservoir $\nu$ is given by 
\begin{equation}
\label{eq:DefinitionHeatNu}
\left<\dot Q^\nu\right> = (\varepsilon-\mu^\nu)\left<I_{\text{M}}^\nu\right>-\alpha V \left<xI_{\text{M}}^\nu\right>. 
\end{equation}
Here, $\left<I_{\text{M}}^\nu\right> \equiv \mathbb E\left[dq^\nu/dt\right]$ is the matter current from reservoir $\nu$,
\begin{equation}
\label{eq:I_avg}
\left<I_{\text{M}}^\nu\right> = \int\!\!\!  dx dv \left[ R^\nu_{10}(x)p(x,v,0,t)-R^\nu_{01}(x)p(x,v,1,t) \right] ,
\end{equation}
and
\begin{equation}
\label{eq:xI_avg}
\left<xI_{\text{M}}^\nu\right> =  \int\!\!\! dx dv x\left[R^\nu_{10}(x)p(x,v,0,t)-R^\nu_{01}(x)p(x,v,1,t)\right]
\end{equation}
represents the position-current correlation.
The average chemical work is given by
\begin{equation}
\label{eq:DefinitionWorkChem}
\left<\dot W^{\text{chem}}\right> = \sum_\nu \mu^\nu \left<I_{\text{M}}^\nu\right>,
\end{equation}
and the heat current entering from the reservoir of the oscillator is\footnote{This can be seen by the connection $v \circ dB = (v + dv / 2) \cdot dB$, where $\cdot$ refers to Itô-type calculus.}
\begin{equation}
\label{eq:DefinitonHeatOsc}
\left<\dot Q^{\text{osc}}\right> = -\gamma \left(\left<v^2\right> - \frac{1}{m\beta^{\text{osc}}}\right).
\end{equation}
The latter equation is formally equivalent to the definition of heat flow for underdamped Langevin dynamics \cite{SekimotoBook2010}. Similarly, the definition of the chemical work flow [see Eq.~(\ref{eq:DefinitionWorkChem})] is consistent with the corresponding definition for the SET (see, e.g., Refs.~\cite{EspositoLindenbergVandenBroeckEPL2009} and \cite{SchallerBook2014} and references therein). However, the definition of heat with respect to left and right leads [see Eq.~(\ref{eq:DefinitionHeatNu})] differs by the additional contribution of $-\alpha V \left<x I_{\text{M}}^\nu\right>$, which stems from the interaction of QD and oscillator. Note that the above definitions of average chemical work flow and average heat flows can also be derived by use of the FPE, Eq.~(\ref{eq:FullFPE}). 

To establish that the second law holds, i.e., that the average total entropy production rate is non-negative, we consider the evolution of the Shannon entropy 
\begin{equation}
\label{eq:S(t)}
S(t)= -\int dxdv\sum\limits_{q}p(x,v,q,t) \ln p(x,v,q,t), 
\end{equation}
where $p(x,v,q,t)$ is the solution of the FPE, Eq.~(\ref{eq:FullFPE}). Taking the time derivative of $S(t)$, introducing the shorthand notation $p(q)\equiv p(x,v,q,t)$, $p(q')\equiv p(x,v,q^\prime,t)$, and $\sumint \equiv \int dx \int dv \sum_q$, and using the conservation of probability as well as partial integration (assuming vanishing boundary contributions, $\lim_{x\to\pm\infty} xp = \lim_{v\to\pm\infty} vp = 0$) we obtain
\begin{equation}
\label{eq:StochasticEntropy}
\begin{aligned}
\frac{d}{dt}S(t) =& \sumint \left[\partial_v J(x,v,q,t)\right] \ln p(q) - \sumint \sum\limits_{q'\nu}R_{qq'}^\nu(x) p(q') \ln p(q),
\end{aligned}
\end{equation}
where 
\begin{equation}
J (x,v,q,t) = -\frac{\gamma}{m} v p(q) - D\partial_v p(q)
\end{equation}
is a probability current. 
Letting $\dot S_1(t)$ and $\dot S_2(t)$ denote the two terms on the right side of Eq.~(\ref{eq:StochasticEntropy}), we integrate by parts to rewrite the first term as follows:
\begin{equation}
\label{eq:S1dot}
\dot S_1(t) =  \sumint \left\{ \frac{\gamma}{m} v \partial_v p(q) + D \frac{[\partial_v p(q)]^2}{p(q)} \right\} .
\end{equation}
From Eq.~(\ref{eq:DefinitonHeatOsc}) we obtain
\begin{equation}
\label{eq:0}
0 = \beta^{\text{osc}} \left<\dot Q^{\text{osc}}\right> + \sumint \left[ \beta^{\text{osc}} \gamma v^2 p(q) + \frac{\gamma}{m} v\partial_v p(q) \right] .
\end{equation}
Summing Eqs.~(\ref{eq:S1dot}) and (\ref{eq:0}) 
we arrive at
\begin{equation}
\label{eq:S1dot_final}
\dot S_1(t) = \beta^{\text{osc}} \left<\dot Q^{\text{osc}}\right> + \dot \Sigma_{\text{cont}} ,
\end{equation}
where
\begin{equation}
\label{eq:EntropyProductionContinuous}
 \dot \Sigma_{\text{cont}}  =  \sumint \frac{\left[\gamma v p(q) + Dm\partial_v p(q)\right]^2}{Dm^2p(q)} \geq 0.
\end{equation}


Next, we rewrite the second term on the right side of Eq.~(\ref{eq:StochasticEntropy}) as follows:
\begin{equation}
\label{eq:S2dot}
\dot S_2(t) = -\frac{1}{2} \sumint \sum\limits_{q'\nu}\left[R_{qq'}^\nu p(q') \ln p(q) + R_{q'q}^\nu p(q)\ln p(q')\right] .
\end{equation}
From the property of (local) detailed balance obeyed by the electron tunnelling rates [see Eq.~(\ref{eq:Rate})], i.e.\
\begin{equation}
\label{eq:LocalDetailedBalance}
\frac{R_{01}^\nu}{R_{10}^\nu} = e^{\beta^\nu(\varepsilon - \alpha V x -\mu^\nu)},
\end{equation}
we derive the identity
\begin{equation}
\label{eq:identity}
0 = 
\sum_\nu \beta^\nu \left<\dot Q^\nu\right>  - \frac{1}{2} \sumint \sum\limits_{q'\nu} \left[ R_{qq'}^\nu p(q') - R_{q'q}^\nu p(q) \right] \ln\frac{R_{q'q}^\nu}{R_{qq'}^\nu} , \\
\end{equation}
where the first term on the right relates to heat exchange with the fermionic leads [see Eqs.~(\ref{eq:DefinitionHeatNu}) - (\ref{eq:xI_avg})].
Summing Eqs.~(\ref{eq:S2dot}) and (\ref{eq:identity}) and rearranging terms, we obtain
\begin{equation}
\label{eq:S2dot_final}
\dot S_2(t) = \sum\limits_\nu \beta^\nu \left<\dot Q^\nu\right> + \dot \Sigma_{\text{disc}},
\end{equation}
where
\begin{equation}
\label{eq:EntropyProductionDiscrete}
\dot \Sigma_{\text{disc}} = \frac{1}{2} \sumint \sum\limits_{q'\nu}\left[R_{qq'}^\nu p(q') - R_{q'q}^\nu p(q)\right]\ln\frac{R_{qq'}^\nu p(q')}{R_{q'q}^\nu p(q)}  \geq 0 .
\end{equation}
Here, non-negativity follows from the log-sum inequality.

Adding Eqs.~(\ref{eq:S1dot_final}) and (\ref{eq:S2dot_final}), we find that the rate of change of the system's Shannon entropy is given by
\begin{equation}
\label{eq:EntropyBalanceStochastic}
\frac{d}{dt} S(t) = \dot S_e + \dot \Sigma,
\end{equation}
with 
\begin{align}
\dot S_e &= \beta^{\text{osc}}\left<\dot Q^{\text{osc}}\right> + \sum\limits_\nu \beta^\nu \left<\dot Q^\nu\right> \\
\label{eq:EntropyProductionStochastic}
\dot \Sigma &= \dot \Sigma_{\text{cont}} + \dot \Sigma_{\text{disc}} \geq 0 .
\end{align}
Here, the entropy flow rate $\dot S_e$ is the rate at which the total entropy of the reservoirs {\it decreases} due to heat exchange with the system \cite{VanDenBroeckPCC2013}.
The quantity
\begin{equation}
\label{eq:Sigmadot}
\dot \Sigma = \frac{d}{dt}  S - \beta^{\text{osc}}\left<\dot Q^{\text{osc}}\right> - \sum\limits_\nu \beta^\nu \left<\dot Q^\nu\right>
\end{equation}
is the total entropy production rate, which can be expressed as the sum of two independently non-negative contributions [Eq.~\ref{eq:EntropyProductionStochastic})], from the \textit{continuous} [Eq.~(\ref{eq:EntropyProductionContinuous})] and \textit{discrete} [Eq.~(\ref{eq:EntropyProductionDiscrete})] degrees of freedom.
The non-negativity of $\dot\Sigma$, Eq.~(\ref{eq:EntropyProductionStochastic}), shows that the second law holds in our system. 

We note that the two separate parts of the total entropy production rate [see Eqs.~(\ref{eq:EntropyProductionContinuous}) and (\ref{eq:EntropyProductionDiscrete})] are formally equivalent to the definitions derived for an independent underdamped harmonic oscillator and an independent SET \cite{EspositoLindenbergVandenBroeckEPL2009, SchallerBook2014}. 
However, as is apparent especially at steady states, where $\frac{d}{dt}S(t) = 0$, the entropy production rate involves the heat flows [see Eq.~(\ref{eq:Sigmadot})], for which the definitions for the independent systems differ from the definition for the electron shuttle [see Eqs.~(\ref{eq:DefinitionHeatNu}) and (\ref{eq:DefinitonHeatOsc})].

\subsection{\label{subsec:MeanFieldThermo}Mean-field thermodynamics}
At the mean field level, the total energy of the system (denoted by an overbar) corresponding to Eqs.~(\ref{eq:MeanField1})-(\ref{eq:MeanField3}) is given by 
\begin{equation}
\bar E = \frac{m\bar v^2}{2}+\frac{k\bar x^2}{2}+\varepsilon\bar q - \alpha V \bar x \bar q ,
\end{equation}
and its rate of change is
\begin{equation}
\frac{d \bar E}{dt} = m\bar v \frac{d{\bar v}}{dt} + k\bar x \bar v + \varepsilon \bar I_{\text{M}} -\alpha V \bar x \bar I_{\text{M}} -\alpha V \bar q \bar v, 
\end{equation}
where
\begin{equation}
\label{eq:IMbar}
\bar I_{\text{M}} = \frac{d\bar q}{dt} = \sum_\nu \left( R_{10}^\nu \bar p_0 - R_{01}^\nu \bar p_1 \right) \equiv \sum\limits_\nu \bar I_{\text{M}}^\nu .
\end{equation}
Note that $\bar I_{\text{M}}^\nu>0$ denotes the flow of matter from reservoir $\nu$ into the system, and vice versa for $\bar I_{\text{M}}^\nu<0$.
Using Eqs.~(\ref{eq:MeanField1})-(\ref{eq:MeanField3}), the first law of thermodynamics takes the form
\begin{equation}
\label{eq:FirstLawMF}
\begin{aligned}
\frac{d\bar E}{dt} &= \sum\limits_\nu (\varepsilon-\alpha V\bar x-\mu^\nu)\bar I_{\text{M}}^\nu +\mu^\nu \bar I_{\text{M}}^\nu -\gamma \bar v^2 \\
&=\dot{\bar{Q}}^{\text{L}} + \dot{\bar{Q}}^{\text{R}} + \dot{\bar{W}}_{\text{chem}}+\dot{\bar{Q}}^{\text{osc}}.
\end{aligned}
\end{equation}
Here
\begin{equation}
\label{eq:DefinitionHeatMF}
\dot{\bar{Q}}^\nu = (\varepsilon-\alpha V\bar x-\mu^\nu)\bar I_{\text{M}}^\nu
\end{equation}
is the heat flow into the QD from reservoir $\nu$,
\begin{equation}
\label{eq:ChemicalWorkMF}
\dot{\bar{W}}^{\text{chem}} = \sum_\nu \mu^\nu \bar I_{\text{M}}^\nu
\end{equation}
is the rate at which chemical work is performed on the system, and
\begin{equation}
\label{eq:DefintionHeatOscMF}
\dot{\bar{Q}}^{\text{osc}} = -\gamma \bar v^2
\end{equation}
is the heat flow into the oscillator from its bath.
Note that $\dot{\bar{Q}}^{\text{osc}}$ is always negative, in contrast to the stochastic case [see Eq.~(\ref{eq:DefinitonHeatOsc})]. 

To establish the second law within the MF approximation, we must first define the system entropy at this level of description.
In the MF equations of motion, the state of the harmonic oscillator $(\bar x,\bar v)$ evolves deterministically under Eqs.~(\ref{eq:MeanField1})-(\ref{eq:MeanField2}), while the quantum dot is represented by a probability distribution $\bar\pb = (\bar p_0,\bar p_1)^T$ evolving under a master equation, Eq.~(\ref{eq:MeanField3}).
We therefore define the entropy of the system to be the Shannon entropy of the QD probability distribution:
\begin{equation}
\label{eq:Sbardef}
\bar S(t) = -\sum_q \bar p_q(t) \ln \bar p_q(t).
\end{equation}
As in Sec.~\ref{subsec:StochasticThermodynamics}, the total entropy production rate is the sum of the rates of change of the entropies of the system and the reservoirs:
\begin{equation}
\label{eq:SigmadotMF}
\dot{\bar \Sigma} = \frac{d}{dt}{\bar{S}}-\beta^{\text{osc}}\dot{\bar Q}^{\text{osc}}-\sum\limits_\nu \beta^\nu \dot{\bar{Q}}^\nu .
\end{equation}
We now analyze the three terms on the right side of this equation.

The rate of change of the system entropy is given by
\begin{equation}
\begin{aligned}
\frac{d}{dt}{\bar S} &= - \sum_\nu \sum_{q,q'} R_{qq'}^\nu \bar p_{q'} \ln \bar p_q \\
&= \sum_\nu \left( R_{10}^\nu \bar p_0 - R_{01}^\nu \bar p_1 \right) \ln \frac{\bar p_0}{\bar p_1} .
\end{aligned}
\end{equation}
By Eq.~(\ref{eq:DefintionHeatOscMF}), the second term on the right side of Eq.~(\ref{eq:SigmadotMF}) is equal to $\gamma\beta^{\text{osc}}\bar v^2$.
To analyze the third term we use Eqs.~(\ref{eq:LocalDetailedBalance}), (\ref{eq:IMbar}) and (\ref{eq:DefinitionHeatMF}) to write
\begin{equation}
\begin{aligned}
\beta^\nu \dot{\bar{Q}}^\nu &= \beta^\nu (\varepsilon-\alpha V\bar x-\mu^\nu)\bar I_{\text{M}}^\nu \\
&= \left( R_{10}^\nu \bar p_0 - R_{01}^\nu \bar p_1 \right) \ln \frac{R_{01}^\nu}{R_{10}^\nu}.
\end{aligned}
\end{equation}
Combining results, we obtain
\begin{equation}
\label{eq:EntropyBalanceMF}
\dot{\bar \Sigma} = \sum_\nu \left( R_{10}^\nu \bar p_0 - R_{01}^\nu \bar p_1 \right) \ln \frac{R_{10}^\nu \bar p_0}{R_{01}^\nu \bar p_1} + \gamma\beta^{\text{osc}}\bar v^2 \geq 0 ,
\end{equation}
in agreement with the second law.

\subsection{\label{subsec:MSPTTD}Perturbative thermodynamics based on multiple scales}
Multiple scale perturbation theory gives the following result for the shuttle probability density (see Sec.~\ref{subsec:MSPT}):
\begin{equation}
\label{eq:ApproximationProb2}
p(x,v,q,t) \approx \pi_q(x)\tilde p(x,v,t), 
\end{equation}
where $\pi_q(x)$ denotes the instantaneous equilibrium distribution of the QD and $\tilde p(x,v,t)$ is the solution to Eq.~(\ref{eq:FokkerPlanckHO}).
Eq.~(\ref{eq:ApproximationProb2}) represents an approximate solution of the FPE, Eq.~(\ref{eq:FullFPE}).
Therefore, to compute thermodynamic quantities such as heat flows and chemical work at this level of approximation, we use Eq.~(\ref{eq:ApproximationProb2}) to evaluate the relevant averages introduced in Sec.~\ref{subsec:StochasticThermodynamics}.

Note that an alternative attempt, which we will not follow here, could be to derive the laws of thermodynamics based on the effective FPE, Eq.~(\ref{eq:FokkerPlanckHO}). This effective FPE goes, however, beyond a simple adiabatic approximation and hence, its associated entropy production rate does not match the original entropy production rate evaluated with the approximated solution \cite{EspositoPRE2012}.

To evaluate the heat flow into the oscillator, Eq.~(\ref{eq:DefinitonHeatOsc}), we first write
\begin{equation}
\begin{aligned}
\left<v^2\right> &= \int dxdv\sum\limits_q v^2 p(x,v,q,t) \\
&\approx \int dxdv\, v^2 \tilde p(x,v,t) = \left< v^2\right>_{\text{MS}},
\end{aligned}
\end{equation}
where $\left<\bullet \right>_{\text{MS}}$ denotes an average taken with the density $\tilde p(x,v,t)$.
Using the transformation to energy $\mathcal E$ and oscillation phase $\theta$ given by Eq.~(\ref{eq:Transformation}), and assuming $\hat p(\mathcal E,\theta,t)\approx \hat p(\mathcal E,t)$ (see Sec.~\ref{subsec:MSPT}), we get
\begin{equation}
\begin{aligned}
\left<v^2\right>_{\text{MS}} &= \int d\mathcal E \frac{1}{m}\mathcal E \, \hat p(\mathcal E, t) = \frac{1}{m}\left<\mathcal E\right>_{\text{MS}},
\end{aligned}
\end{equation}
after averaging over $\theta$.
Here, $\hat p(\mathcal E,t)$ is the solution to Eq.~(\ref{eq:FPEEnergy}), which at steady state is given by Eq.~(\ref{eq:SolutionMSPT}). 

For the chemical work and the heat exchanges with fermionic reservoirs, we need the matter currents $\left<I_{\text{M}}^\nu\right>$ and position-current correlations, $\left<xI_{\text{M}}^\nu\right>$.
Substituting Eq.~(\ref{eq:ApproximationProb2}) into Eq.~(\ref{eq:I_avg}) we get
\begin{equation}
\left<I_{\text{M}}^\nu\right> \approx  \left<\left[R_{10}^\nu(x)\pi_0(x) -R_{01}^\nu(x)\pi_1(x)\right]\right>_{\text{MS}}.
\end{equation}
Transforming to $(\mathcal E,\theta)$-space and averaging over $\theta$ gives
\begin{equation}
\left<I_{\text{M}}^\nu\right>_{\text{MS}} = \int d\mathcal E \hat R^\nu(\mathcal E)\hat p(\mathcal E,t)  = \left<\hat R^\nu(\mathcal E)\right>_{\text{MS}}
\end{equation}
with 
\begin{equation}
\hat R^\nu(\mathcal E) = \frac{1}{2\pi}\int\limits_0^{2\pi} \left[R_{10}^\nu(x)\pi_0(x) -R_{01}^\nu(x)\pi_1(x)\right] d\theta.
\end{equation}
For the position-current correlations, we similarly get
\begin{equation}
\left<xI_{\text{M}}^\nu\right> \approx \left<\reallywidehat{x R}^\nu(\mathcal E)\right>_{\text{MS}}
\end{equation}
with 
\begin{equation}
\reallywidehat{x R}^\nu(\mathcal E) = \frac{1}{2\pi}\int\limits_0^{2\pi}x\left[R_{10}^\nu(x)\pi_0(x) -R_{01}^\nu(x)\pi_1(x)\right]d\theta.
\end{equation}
Combining results with Eqs.~(\ref{eq:DefinitionHeatNu}), (\ref{eq:DefinitionWorkChem}) and (\ref{eq:DefinitonHeatOsc}) we obtain
\begin{align}
\label{eq:HeatFlowNuMSPT}
\left<\dot Q^\nu\right>_{\text{MS}} &= \left(\varepsilon-\mu^\nu\right)\left<\hat R^\nu\right>_{\text{MS}}-\alpha V \left<\reallywidehat{xR}^\nu\right>_{\text{MS}}, \\
\label{eq:chemicalWorkMSPT}
\left<\dot W^{\text{chem}}\right>_{\text{MS}} &= \sum\limits_\nu \mu^\nu \left<\hat R^\nu\right>_{\text{MS}}, \\
\label{eq:HeatFlowOscMSPT}
\left<\dot Q^{\text{osc}}\right>_{\text{MS}} &= -\frac{\gamma}{m}\left(\left<\mathcal E\right>_{\text{MS}}-\frac{1}{\beta^{\text{osc}}}\right).
\end{align}
As in Sec.~\ref{subsec:StochasticThermodynamics}, the first law is expressed by Eq.~(\ref{eq:AverageFirstLaw}), but the heat flows and chemical work are now given by Eqs.~(\ref{eq:HeatFlowNuMSPT})-(\ref{eq:HeatFlowOscMSPT}).

Using Eq.~(\ref{eq:ApproximationProb2}), the system entropy [Eq.~(\ref{eq:S(t)})] becomes a sum of distinct contributions from the harmonic oscillator and the quantum dot:
\begin{equation}
S(t) \approx \left< -\ln \tilde p + S_{QD}\right>_{\text{MS}}
\end{equation}
where $S_{QD}(x) = -\sum_q \pi_q(x) \ln \pi_q(x)$.

The decomposition $\dot S = \dot S_e + \dot \Sigma$ [see Eq.~(\ref{eq:EntropyBalanceStochastic})] remains valid in the MS approximation.
To show that the entropy production rate $\dot\Sigma$ is non-negative at this level of approximation, we first look at the contribution from the continuous degrees of freedom.
Replacing $p(x,v,q,t)$ by $\pi_q(x)\tilde p(x,v,t)$ in Eq.~(\ref{eq:EntropyProductionContinuous}), we find
\begin{equation}
\begin{aligned}
\dot \Sigma_{\text{cont}} &\approx \int dxdv \sum\limits_{q} \frac{\left[\gamma v \pi_q \tilde p +Dm \partial_v(\pi_q \tilde p)\right]^2}{Dm^2 \pi_q\tilde p}\\
&= \int dxdv \frac{\left[\gamma v  \tilde p+Dm \partial_v \tilde p\right]^2}{Dm^2 \tilde p}\geq 0.
\end{aligned}
\end{equation}
Transforming to $(\mathcal E,\theta)$-space and averaging over $\theta$ then results in 
\begin{equation}
\begin{aligned}
\dot \Sigma_{\text{cont,MS}} &= \int d\mathcal E \frac{\left[\gamma\sqrt{\mathcal E}\hat p+D\sqrt{\mathcal E}m^2\partial_{\mathcal E} \hat p\right]^2}{Dm^3\hat p}\\
&\geq 0.
\end{aligned}
\end{equation}
Applying a similar analysis for the discrete degrees of freedom [see Eq.~(\ref{eq:EntropyProductionDiscrete})] we get
\begin{equation}
\dot \Sigma_{\text{disc}}\approx \frac{1}{2} \sumint \sum\limits_{q'\nu}\left(R_{qq'}^\nu\pi_{q'}-R_{q'q}^\nu\pi_q\right)\ln\frac{R_{qq'}^\nu \pi_{q'}}{R_{q'q}^\nu \pi_{q}}\tilde p,
\end{equation}
which after transforming to $(\mathcal E,\theta)$-space and averaging over $\theta$ becomes
\begin{equation}
\dot \Sigma_{\text{disc,MS}} = \frac{1}{2}\int d\mathcal E~\sigma(\mathcal E)\hat p(\mathcal E,t) \geq 0, 
\end{equation}
where 
\begin{equation}
\sigma(\mathcal E) = \frac{1}{2\pi}\int\limits_0^{2\pi}d\theta \sum\limits_{qq'\nu}\left(R_{qq'}^\nu\pi_{q'}-R_{q'q}^\nu\pi_q\right)\ln\frac{R_{qq'}^\nu \pi_{q'}}{R_{q'q}^\nu \pi_{q}}.
\end{equation}
$\dot \Sigma_{\text{disc,MS}}$ is non-negative since $\sigma(\mathcal E) \geq 0$ by the log-sum inequality, and $\hat p(\mathcal E,t)\geq 0$ is a probability density.
Combining results, we get
\begin{equation}
\label{eq:EntropyProductionMSPT}
\dot\Sigma_{\text{MS}}=\dot\Sigma_{\text{cont,MS}}+\dot\Sigma_{\text{disc,MS}}\geq 0,
\end{equation}
again in agreement with the second law.

\subsection{\label{subsec:DiscussionTD}Discussion}
\begin{figure}[h]
\begin{center}
\includegraphics[width=0.7\columnwidth]{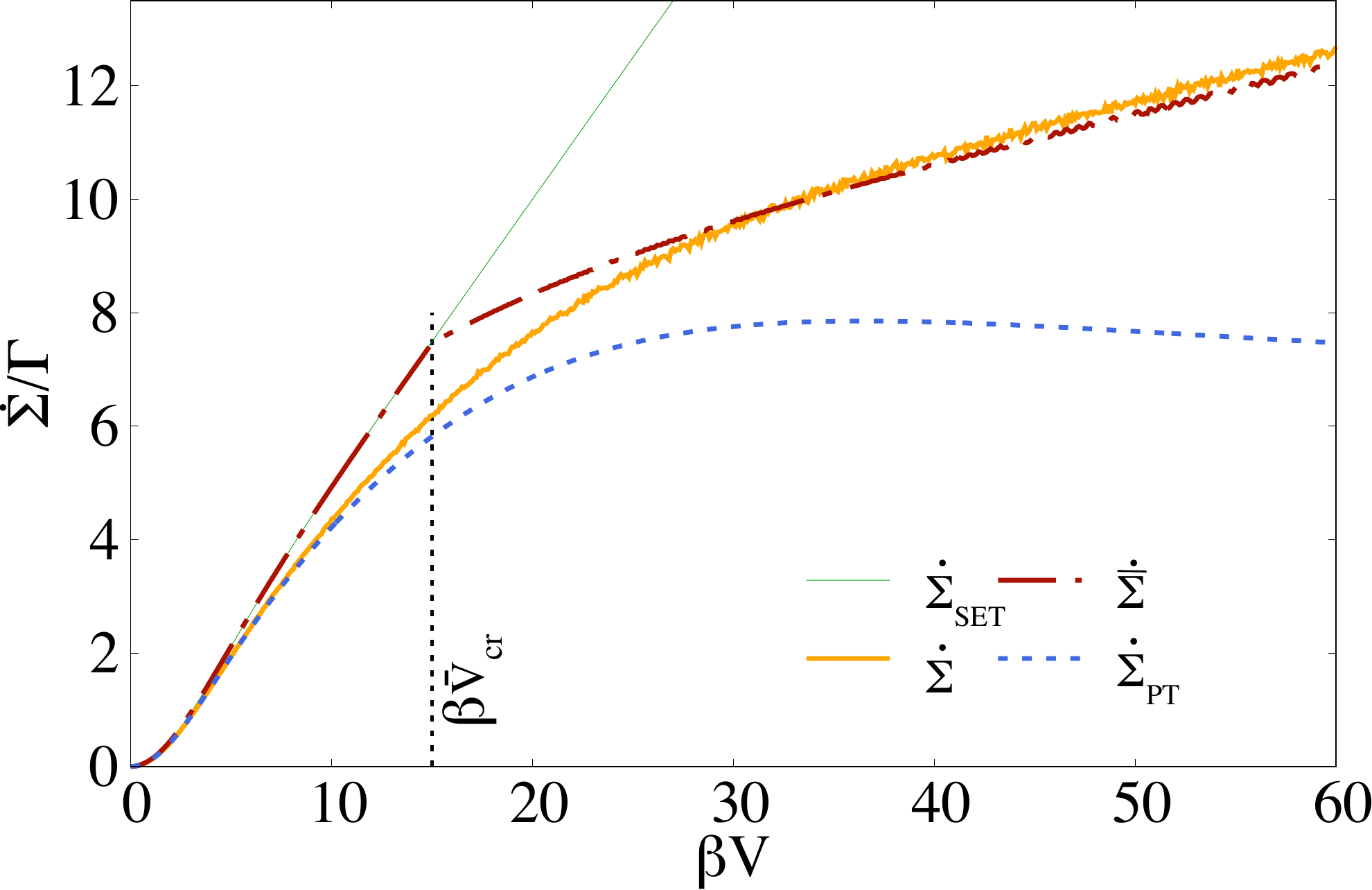}
\end{center}
\caption{
Entropy production rate for the full stochastic description [Eq.~(\ref{eq:EntropyProductionStochastic}), orange solid], the MF approximation [Eq.~(\ref{eq:EntropyBalanceMF}), red dash-dotted], the MS analysis [Eq.~(\ref{eq:EntropyProductionMSPT}), blue dotted)], and for the SET [thin green, Eq.~(\ref{eq:EntropyProductionDiscrete}) with $\alpha\to 0 \, , \, \lambda\to\infty$], as a function of bias voltage.
The MF entropy production shows a clear signature of the underlying bifurcation at the critical value $\beta \bar V_{\text{cr}} = 15.0$.
The MS entropy production rate deviates from (specifically, underestimates) the full entropy production rate as $V$ is increased, due to the breakdown of time scale separation.}
\label{fig:Entropy_beta_1}
\end{figure}

Having derived the general laws of thermodynamics at the different levels, we now discuss the thermodynamic properties of the model at hand and compare the stochastic, MF and MS solutions, focusing on the transition towards self-oscillation. All numerical results are obtained using the parameters specified in \ref{secApp:ComputationalMethods}.

\subsubsection{Entropy production rate}
We first look at the steady state entropy production rates as a function of the applied bias voltage $V$. Fig.~\ref{fig:Entropy_beta_1} shows the entropy 
production rate for the stochastic case (orange solid), the MF case (red dash-dotted) and the MS description (blue dotted).
Also shown is the single electron transistor entropy production rate $\dot \Sigma_{\text{SET}}$ (green thin), 
i.e. Eq.~(\ref{eq:EntropyProductionDiscrete}) for the case where the position $x$ is fixed at the origin 
\cite{SchallerBook2014, EspositoLindenbergVandenBroeckEPL2009} or, alternatively, in the completely decoupled limit $\alpha\to 0 \, , \, \lambda\to\infty$ (see Sec.~\ref{subsec:Stochastic}).

The steady state entropy production rate at the MF and the stochastic level is equal to the chemical work, aside from a factor $\beta$.
To see this in the stochastic case, note that in the steady state the system's Shannon entropy and average energy are constant: $dS/dt = 0 = \langle dE/dt \rangle$.
Combining this observation with Eqs.~(\ref{eq:AverageFirstLaw}), (\ref{eq:DefinitionWorkChem}) and (\ref{eq:Sigmadot}), and with our choice of setting all reservoir temperatures to be equal, $\beta^\nu = \beta^{\text{osc}} = \beta$, we obtain
\begin{equation}
\label{eq:EntropyAsWork}
\dot \Sigma = \beta \left<\dot W^{\text{chem}}\right> = \beta V\left<I_{\text{M}}^{\text{L}}\right> = -\beta V\left<I_{\text{M}}^{\text{R}}\right>.
\end{equation}
The last equality follows from the preservation of electron number, $\left<I_{\text{M}}^{\text{R}}\right> +\left<I_{\text{M}}^{\text{L}}\right> = 0$.
In the MF case, we arrive at the analogous result using Eqs.~(\ref{eq:FirstLawMF}), (\ref{eq:ChemicalWorkMF}) and (\ref{eq:SigmadotMF}).

At the MF level, we see in Fig.~\ref{fig:Entropy_beta_1} that below $\beta \bar V_{\text{cr}}=15.0$ the entropy production rate $\dot{\bar \Sigma}$ is essentially the same as for the SET. 
This is understandable: below the bifurcation, the system evolves to a stable fixed point, with the quantum dot at rest near the origin (see Sec.~\ref{subsec:mfd}).
Above the bifurcation the shuttle oscillates, hence the resulting entropy production deviates from that of the SET.
The sharp transition to self-oscillation is clearly reflected in the deviation of $\dot{\bar \Sigma}$ from $\dot \Sigma_{\text{SET}}$ for $V>\bar V_{\text{cr}}$.

Interestingly, we see in Fig.~\ref{fig:Entropy_beta_1} that self-oscillation lowers the rate of entropy production, relative to the value it would have taken had the quantum dot remained at rest; that is, $\dot{\bar \Sigma} < \dot \Sigma_{\text{SET}}$ for $V>\bar V_{\text{cr}}$.
In effect, above the critical voltage, when faced with a ``choice'' between two modes of behavior -- oscillatory or at rest -- the shuttle adopts the one that generates entropy more slowly. To understand this point quantitatively, note that the entropy production rate is determined by the matter current flowing from left to right through the device, Eq.~(\ref{eq:EntropyAsWork}).
For $\beta V \gg 1$ the SET current approaches $I_{\text{M}}^{\text{SET}} = \Gamma/2$, as our choice of chemical potentials produces a SET steady state in which $p_0=p_1=1/2$.
For the MF case, recall that for $V\gg \bar V_{\text{cr}}$ our system approaches the perfect shuttling regime in which one electron is transferred per oscillation period, which implies $\bar I_{\text{M}} = \omega/2\pi$, where $\omega$ is the oscillation frequency.
Our parameter choices give $I_{\text{M}}^{\text{SET}} = 0.5$ and $\bar I_{\text{M}} \approx 0.1$, hence the SET generates entropy at a rate about five times that of the MF shuttle, in the high-bias limit. 
These results are consistent with the asymptotic slopes of the SET and MF entropy production rates shown in Fig.~\ref{fig:Entropy_beta_1}.
By the same token, if the parameters were chosen such that $\bar I_{\text{M}} > I_{\text{M}}^{\text{SET}}$ (roughly, if $\omega =\sqrt{k/m} > \pi\Gamma$), then we would get $\dot{\bar \Sigma} > \dot \Sigma_{\text{SET}}$.

In the stochastic case the oscillator undergoes thermal motions in the steady state, the matter current through the quantum dot is lower than the corresponding SET current, and as a result $\dot \Sigma < \dot \Sigma_{\text{SET}}$. 
Note that the onset of oscillations is not as clearly marked as in the MF case, rather the entropy production rate transitions smoothly from one regime to the other.From Fig.~\ref{fig:Entropy_beta_1} we see that for both small and large bias voltages, the MF and stochastic entropy production rates agree quite well: both approach the SET value when $V\ll \bar V_{\text{cr}}$, and when $V\gg \bar V_{\text{cr}}$ the MF value only slightly underestimates the entropy production rate.
However, around the bifurcation at $\bar V_{\text{cr}}$ the stochastic entropy production rate deviates substantially from the MF prediction. Here, the system shows bistable behaviour, that is, it jumps between oscillating and resting state. Therefore, fluctuations are not small and cannot be neglected, i.e., the MF assumption is no longer valid. Similar behaviour has been observed for the dissipated work of a network of units, for which a sharp transition of a MF bifurcation is smoothed out for small system sizes and the sharp MF transition is only recovered for large network sizes \cite{HerpichEtAlPRX2018}.

We finally note that the stochastic entropy production rate is well approximated by the multi-scale (MS) results, up to $V\approx \bar V_{\text{cr}}$.
As argued in Sec.~\ref{subsec:Dynamics}, for $V > \bar V_{\text{cr}}$ the key assumption of time scale separation is no longer valid and the perturbative solution breaks down.
This is seen in the large deviation of $\dot \Sigma_{\text{MS}}$ from $\dot \Sigma$ in Fig.~\ref{fig:Entropy_beta_1}. 

\begin{figure}[h]
\begin{center}
\includegraphics[width=0.7\columnwidth]{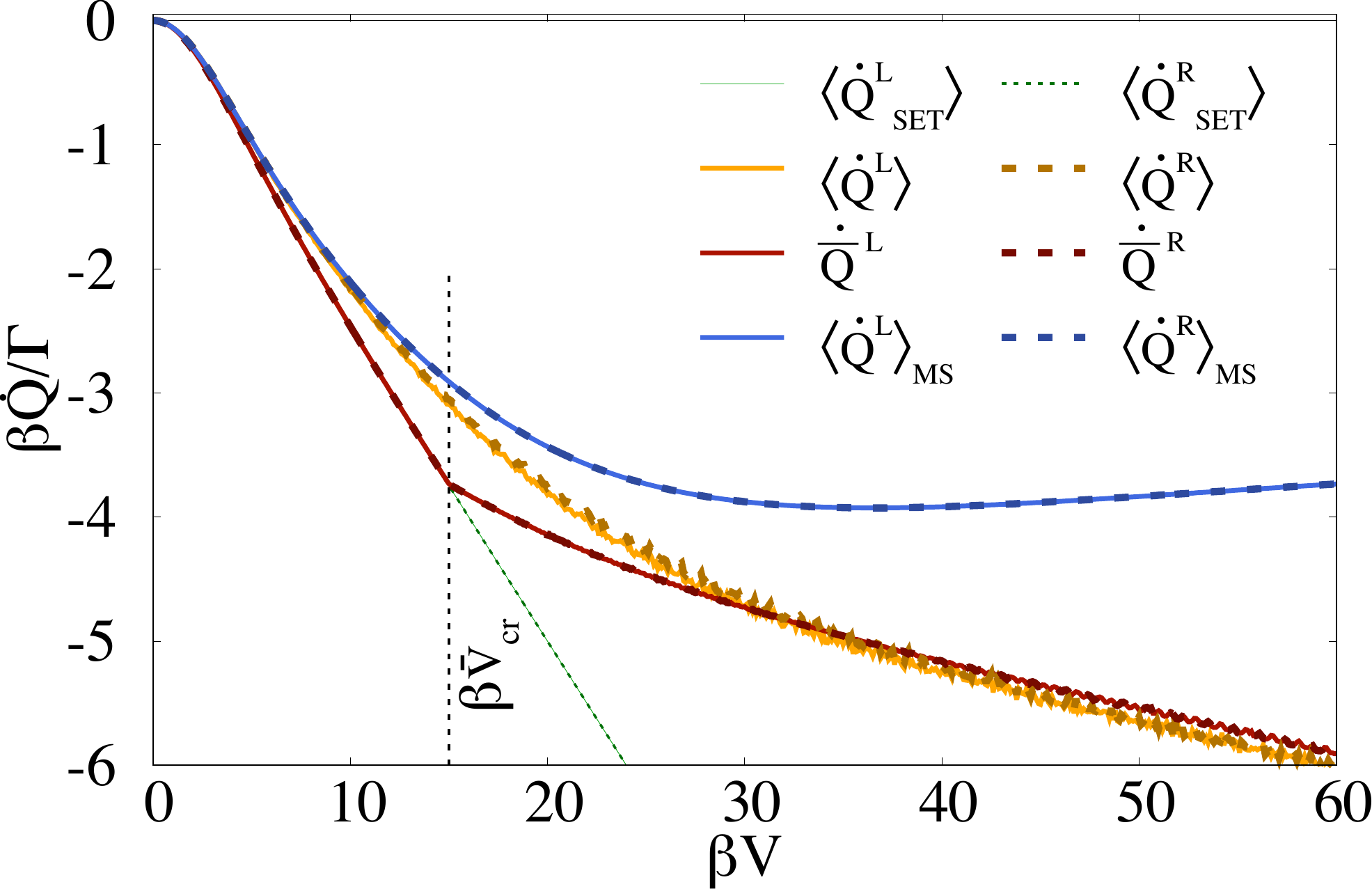}
\end{center}
\caption{Heat flows between the fermionic reservoirs and the QD as a function of the bias voltage $V$: Solid lines correspond to heat flow of the left reservoir, dotted lines to heat flow of the right reservoir. We plot the full stochastic heat flows [orange, Eq.~(\ref{eq:DefinitionHeatNu}), the MF heat flows [red, Eq.~(\ref{eq:DefinitionHeatMF})] as well as the heat flows derived by MS perturbation theory [blue, Eq.~(\ref{eq:HeatFlowNuMSPT})]. For comparison we also plot the heat flows of the SET (thin green). The MF heat flows show a clear signature of the underlying bifurcation whereas the full stochastic and the MS heat flows transition smoothly between the two regimes of operation.}
\label{fig:Thermo_beta_1}
\end{figure}

\begin{figure}[h]
\begin{center}
\includegraphics[width=0.7\columnwidth]{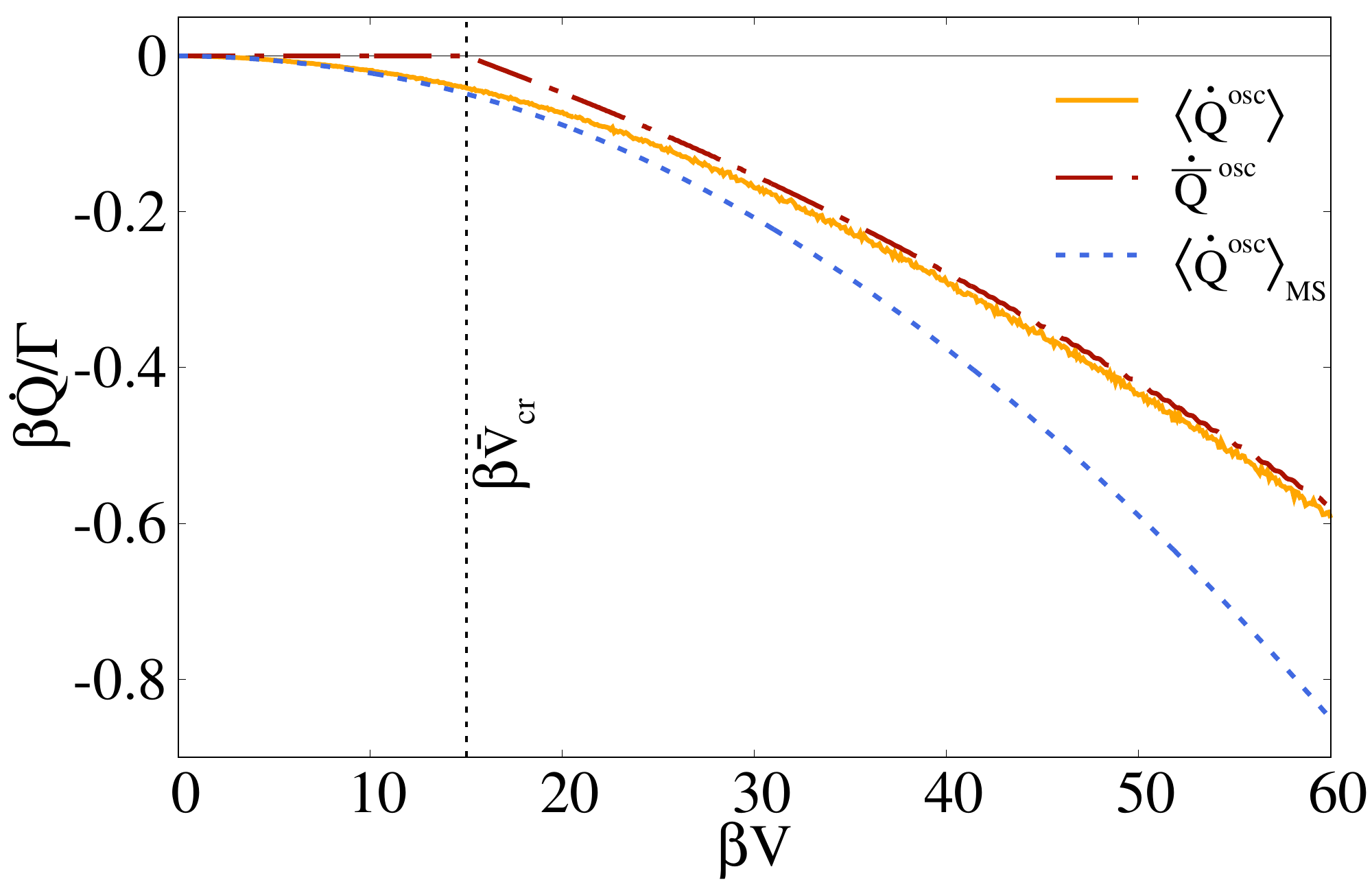}
\end{center}
\caption{Heat flow between the oscillator and its respective heat bath as a function of the bias voltage $V$: The solid (green) line corresponds to fully stochastic heat flow [see Eq.~\ref{eq:DefinitonHeatOsc}), the (red) dash-dotted line shows the MF heat flow [see Eq.~(\ref{eq:DefintionHeatOscMF})] and the dotted (blue) line shows the effective heat flow by use of MS [see Eq.~(\ref{eq:HeatFlowOscMSPT})]. Again the MF heat flow shows a clear signature of the underlying bifurcation whereas the full stochastic and the MS heat flow transition smoothly between the two regimes of operation.}
\label{fig:Thermo_beta_2}
\end{figure}

\subsubsection{Heat flows}
Next we look at the heat flows between the shuttle and the three reservoirs at steady state. At steady state the average energy of the system is constant at all levels, i.e., $\left<dE/dt\right>  =d\bar E/dt = \left<dE/dt\right>_{\text{MS}}=0$, as we have verified numerically. In Fig.~\ref{fig:Thermo_beta_1} we show the left and right steady state heat flows. At all three levels of description the heat flows between the QD and the left and right reservoirs are nearly indistinguishable.
This is a consequence of our parameter choices of small $\alpha$ and $\mu^\nu = \varepsilon\pm V/2$, as can be seen from Eq.~(\ref{eq:DefinitionHeatNu}): $\left<\dot Q^\nu\right> = (\varepsilon - \mu^\nu)\left<I_{\text{M}}^\nu\right> - \alpha V \left<xI_{\text{M}}^\nu\right>$. 
By the conservation of the matter, the first term on the right is the same for the left and the right reservoir and differences arise only from the correlation of $x$ and $I_{\text{M}}^\nu$. Since $\alpha = 0.06 \ll 1$, the difference is barely noticeable in Fig.~\ref{fig:Thermo_beta_1}.
Note that all heat flows are negative while the chemical work is positive, indicating that chemical work is performed on the system, and energy is transferred as heat into all three reservoirs.

As with the case of entropy production (Fig.~\ref{fig:Entropy_beta_1}), at the MF level the onset of oscillations at $\beta \bar V_{\text{cr}}$ is clearly reflected in the heat flows $\dot{\bar Q}^\nu$ (Fig.~\ref{fig:Thermo_beta_1}) and  $\dot{\bar Q}^{\text{osc}}$ (Fig.~\ref{fig:Thermo_beta_2}).
The latter vanishes below the bifurcation (where the shuttle is at rest), but becomes negative above the bifurcation, where the shuttle's oscillatory motion gives rise to dissipation due to friction.
Below the bifurcation, the MF heat flows agree with the corresponding SET values, as expected.

In the stochastic case the transition to self-oscillation is gradual rather than sharp, as seen in the behaviours of $\left<\dot{Q}^{\text{L}}\right>$, $\left<\dot{Q}^{\text{R}}\right>$ and $\left<\dot{Q}^{\text{osc}}\right>$.
Away from the transition -- that is, for small and large values of $V$ -- these heat flows are well approximated by the MF values, as was the case for the entropy production rate.

For the MS solution we see that $\left<\dot Q^{\nu/\text{osc}}\right> \approx\left<\dot Q^{\nu/\text{osc}}\right>_{\text{MS}}$ at small values of $V$.
At larger values of the applied bias voltage the MS heat flows deviate from the full stochastic heat flows, due to the breakdown of time scale separation as previously discussed.

To summarize this section, the thermodynamic quantities we have studied -- the entropy production rate and the heat exchanges with reservoirs -- all bear the signature of the onset of self-oscillation.
In particular, all of these quantities deviate substantially from the corresponding SET values above the critical voltage $\bar V_{\text{cr}}$, as the system approaches the pure shuttling regime and its oscillatory motion influences the exchange of energy and matter.
The transition is abrupt in the mean field approximation, but gradual in the fully stochastic case.

\section{\label{sec:Conclusion}Conclusion and future applications}
We have provided a classical stochastic description of the single electron shuttle based on coupled Langevin equations 
including thermal and Poissonian noise terms. The average dynamics can be well approximated by MF equations away from the onset of self-oscillations for 
our specific choice of parameters. However, we expect deviations between the stochastic and MF description if fluctuations are strong, i.e. for a system with a smaller mass. Within the MF approximation the system undergoes a Hopf bifurcation from a stable fixed 
point to a stable limit cycle by changing the applied bias voltage, where a limit cycle corresponds to self-oscillation 
of the shuttle. 

By introducing a time scale separation between the frequent tunnelling events and the slow dynamics of the oscillator we were able to perturbatively solve the FPE corresponding to the coupled Langevin equations. This MS solution approximates the full stochastic solution very well below and around the critical bias voltage of the MF description. However, as the applied bias voltage is increased and the system starts shuttling, the underlying assumption of many jumps per cycle is not valid anymore and the perturbation approach breaks down. An order parameter defined in terms of a mean amplitude of oscillation shows clear signatures of the Hopf bifurcation found in the deterministic MF description. This order parameter is not sensitive to small oscillations of the shuttle in the stochastic and MS approaches, hence the system transitions smoothly towards self-oscillation in those cases, and the abrupt onset is realized only in the MF case.

In the classical description chosen here, we identify three different regimes: Below the bifurcation, i.e., for $V< \bar V_{\text{cr}}$ the shuttle acts as a single electron transistor with additional noise. For very large bias voltages $V\gg \bar V_{\text{cr}}$ the system oscillates and transports one electron from the reservoir with higher chemical potential to the reservoir with lower chemical potential per period. In this regime the system truly serves as a shuttle for single electron transport. Between the two limits there is a crossover regime.

Additionally, we performed a thermodynamic analysis of the shuttling mechanism. Using stochastic thermodynamics we derived the first and second law at the stochastic as well as the MF level. At the perturbative level, we used the solution from MS perturbation theory to perform ensemble averages in order to define effective thermodynamic quantities. The thermodynamic quantities as entropy production rate, heat flows and chemical work rate show clear signatures of the underlying bifurcation within the MF approximation. 
The corresponding stochastic and MS quantities lack such an abrupt transition from noisy movement to self-oscillation, but show a smoothed transition, which suggest that the abrupt transition seen in the dynamical description is an artefact of the chosen order parameter.  However, the thermodynamic quantities do reflect the transition towards pure shuttling if compared to the single electron transistor. 

The thermodynamic analysis of the electron shuttle provides an exemplary discussion of the thermodynamics of self-sustained oscillations, especially for highly fluctuating systems at the nano-scale, and which is also experimentally realizable. The combined system of QD and oscillator has rich dynamics and is capable of realizing various thermodynamic objectives. Our thermodynamic analysis from different perspectives may be used as the starting point to analyze the performance of the shuttle as a heat pump or refrigerator, with applied chemical work used to transfer heat from cold fermionic reservoirs to the hot reservoir of the oscillator ($\beta^\nu>\beta^{\text{osc}}$). A quantum heat engine using the electron shuttle has recently been discussed in \cite{TonekaboniEtAlArXiv2018}. Alternatively, the shuttle can be transformed into an engine, which uses the chemical gradient in order to perform mechanical work. Such an engine might be constructed as a nanoscale rotor driven by single electron tunnelling, as proposed in \cite{CroyEisfeldEPL2012} and \cite{ WaechtlerEtAlArXiv2019}. Similarly, one can then use the mechanical motion in order to pump electrons and generate a matter current against an externally applied electric field \cite{WaechtlerEtAlArXiv2019}. The thermodynamic analysis of such a device is the subject of our further research in this direction. With the present work, we hope to pave the way for proper thermodynamic analyses for realistic engines based on the concept of the electron shuttle and stimulate further discussion about the thermodynamic possibilities of such devices. 

\section*{Acknowledgments}
The authors thank T. Brandes for initiating this project. C. W. acknowledges fruitful discussions with S. Restrepo and J. Cerrillo. The authors are thankful for stimulating discussions with H. Engel. This work has been funded by the Deutsche Forschungsgemeinschaft (DFG, German Research Foundation) - Projektnummer 163436311 - SFB 910 and the Graduate Research Training Group RTG 1558. P. S. acknowledges financial support by the European Research Council project Nano Thermo (ERC-2015-CoG Agreement No. 681456) and from the DFG through project STR 1505/2-1. C. J. was supported by the US National Science Foundation under grant DMR-1506969.


\section*{References}

\bibliographystyle{iopart-num}
\bibliography{books,general_QM,info_thermo,mathematical,open_systems,thermo,electronShuttle}

\providecommand{\newblock}{}
\begin{thebibliography}{10}
\expandafter\ifx\csname url\endcsname\relax
  \def\url#1{{\tt #1}}\fi
\expandafter\ifx\csname urlprefix\endcsname\relax\def\urlprefix{URL }\fi
\providecommand{\eprint}[2][]{\url{#2}}

\bibitem{JenkinsPR2013}
Jenkins A 2013 {\em Phys. Rep.\/} {\bf 525} 167--222

\bibitem{VanDerPolVanDerMarkJoS1928}
Van Der~Pol B and Van Der~Mark J 1928 {\em Phil. Mag. Series 7\/} {\bf 6}
  763--775

\bibitem{NovakTysonNature2008}
Nov{\'a}k B and Tyson J~J 2008 {\em Nat. Rev. Mol. Cell Biol.\/} {\bf 9} 981

\bibitem{FieldNoyesJoCP1974}
Field R~J and Noyes R~M 1974 {\em J. Chem. Phys.\/} {\bf 60} 1877--1884

\bibitem{GorelikEtAlPRL1998}
Gorelik L, Isacsson A, Voinova M, Kasemo B, Shekhter R and Jonson M 1998 {\em
  Phys. Rev. Lett.\/} {\bf 80} 4526

\bibitem{BoeseSchoellerEPL2001}
Boese D and Schoeller H 2001 {\em Europhys. Lett.\/} {\bf 54} 668

\bibitem{ArmourMacKinnonPRB2002}
Armour A and MacKinnon A 2002 {\em Phys. Rev. B\/} {\bf 66} 035333

\bibitem{MccarthyEtAlPRB2003}
McCarthy K~D, Prokof'ev N and Tuominen M~T 2003 {\em Phys. Rev. B\/} {\bf 67}
  245415

\bibitem{NovotnyEtAlPRL2003}
Novotn{\`y} T, Donarini A and Jauho A~P 2003 {\em Phys. Rev. Lett.\/} {\bf 90}
  256801

\bibitem{NovotnyEtAlPRL2004}
Novotn{\`y} T, Donarini A, Flindt C and Jauho A~P 2004 {\em Phys. Rev. Lett.\/}
  {\bf 92} 248302

\bibitem{DonariniEtAlNJP2005}
Donarini A, Novotn{\`y} T and Jauho A~P 2005 {\em New J. Phys.\/} {\bf 7} 237

\bibitem{UtamiEtAlPRB2006}
Utami D~W, Goan H~S, Holmes C and Milburn G 2006 {\em Phys. Rev. B\/} {\bf 74}
  014303

\bibitem{FedoretsEtAlEPL2002}
Fedorets D, Gorelik L, Shekhter R and Jonson M 2002 {\em Europhys. Lett.\/}
  {\bf 58} 99

\bibitem{NoceraEtAlPRB2011}
Nocera A, Perroni C, Ramaglia V~M and Cataudella V 2011 {\em Phys. Rev. B\/}
  {\bf 83} 115420

\bibitem{IsacssonEtAlPhysicaB1998}
Isacsson A, Gorelik L, Voinova M, Kasemo B, Shekhter R and Jonson M 1998 {\em
  Physica B\/} {\bf 255} 150--163

\bibitem{WeissZwergerEPL1999}
Weiss C and Zwerger W 1999 {\em Europhys. Lett.\/} {\bf 47} 97

\bibitem{NordEtAlPRB2002}
Nord T, Gorelik L, Shekhter R and Jonson M 2002 {\em Phys. Rev. B\/} {\bf 65}
  165312

\bibitem{ParkEtAlNature2000}
Park H, Park J, Lim A~K, Anderson E~H, Alivisatos A~P and McEuen P~L 2000 {\em
  Nature\/} {\bf 407} 57

\bibitem{MoskalenkoEtAlPRB2009}
Moskalenko A~V, Gordeev S~N, Koentjoro O~F, Raithby P~R, French R~W, Marken F
  and Savel'ev S~E 2009 {\em Phys. Rev. B\/} {\bf 79} 241403

\bibitem{MoskalenkoEtAlNanotechnology2009}
Moskalenko A, Gordeev S, Koentjoro O, Raithby P, French R, Marken F and
  Savel'ev S 2009 {\em Nanotechnology\/} {\bf 20} 485202

\bibitem{KonigWeigAPL2012}
K{\"o}nig D~R and Weig E~M 2012 {\em Appl. Phys. Lett.\/} {\bf 101} 213111

\bibitem{ScheibleBlickAPL2004}
Scheible D~V and Blick R~H 2004 {\em Appl. Phys. Lett.\/} {\bf 84} 4632--4634

\bibitem{KimEtAlAPL2015}
Kim C, Prada M, Qin H, Kim H~S and Blick R~H 2015 {\em Appl. Phys. Lett.\/}
  {\bf 106} 061909

\bibitem{JoachimEtAlNature2000}
Joachim C, Gimzewski J~K and Aviram A 2000 {\em Nature\/} {\bf 408} 541

\bibitem{ShekhterEtAlJoP2003}
Shekhter R, Galperin Y, Gorelik L~Y, Isacsson A and Jonson M 2003 {\em J. Phys.
  Condens. Matter\/} {\bf 15} R441

\bibitem{GalperinEtAlJoP2007}
Galperin M, Ratner M~A and Nitzan A 2007 {\em J. Phys. Condens. Matter\/} {\bf
  19} 103201

\bibitem{GalperinEtAlScience2008}
Galperin M, Ratner M~A, Nitzan A and Troisi A 2008 {\em Science\/} {\bf 319}
  1056--1060

\bibitem{ShekhterEtAlNano2013}
Shekhter R~I, Gorelik L~Y, Krive I~V, Kiselev M, Parafilo A and Jonson M 2013
  {\em Nanoelectromech. Syst.\/} {\bf 1} 1--25

\bibitem{LaiEtAlFoP2015}
Lai W, Zhang C and Ma Z 2015 {\em Front. Phys.\/} {\bf 10} 59--86

\bibitem{StrogatzBook2018}
Strogatz S~H 2018 {\em Nonlinear dynamics and chaos: with applications to
  physics, biology, chemistry, and engineering\/} (CRC Press)

\bibitem{WaechtlerEtAlArXiv2019}
W{\"a}chtler C, Strasberg P and Schaller G 2019 {\em arXiv preprint
  arXiv:1903.07500\/}

\bibitem{EspositoHarbolaMukamelRMP2009}
Esposito M, Harbola U and Mukamel S 2009 {\em Rev. Mod. Phys.\/} {\bf 81} 1665

\bibitem{SekimotoBook2010}
Sekimoto K 2010 {\em Stochastic Energetics\/} (Berlin Heidelberg: Lect. Notes
  Phys., Springer)

\bibitem{CampisiHaenggiTalknerRMP2011}
Campisi M, H\"anggi P and Talkner P 2011 {\em Rev. Mod. Phys.\/} {\bf 83} 771

\bibitem{JarzynskiAnnuRevCondMat2011}
Jarzynski C 2011 {\em Annu. Rev. Condens. Matter Phys.\/} {\bf 2} 329--351

\bibitem{SeifertRPP2012}
Seifert U 2012 {\em Rep. Prog. Phys.\/} {\bf 75} 126001

\bibitem{SchallerBook2014}
Schaller G 2014 {\em Open Quantum Systems Far from Equilibrium\/} (Cham: Lect.
  Notes Phys., Springer)

\bibitem{VandenBroeckEspositoPhysA2015}
{Van den Broeck} C and Esposito M 2015 {\em Physica (Amsterdam)\/} {\bf 418A}
  6--16

\bibitem{SchmiedlSeifertEPL2007}
Schmiedl T and Seifert U 2007 {\em Europhys. Lett.\/} {\bf 81} 20003

\bibitem{BlickleBechingerNature2011}
Blickle V and Bechinger C 2011 {\em Nat. Phys.\/} {\bf 8} 143

\bibitem{RanaEtAlPRE2014}
Rana S, Pal P~S, Saha A and Jayannavar A~M 2014 {\em Phys. Rev. E\/} {\bf
  90}(4) 042146

\bibitem{HolubecJSM2014}
Holubec V 2014 {\em J. Stat. Mech. Theor. Exp.\/} {\bf 2014} P05022

\bibitem{MartinezNature2016}
Mart{\'\i}nez I~A, Rold{\'a}n {\'E}, Dinis L, Petrov D, Parrondo J~M and Rica
  R~A 2016 {\em Nat. Phys.\/} {\bf 12} 67

\bibitem{RestrepoEtAlNJP2018}
Restrepo S, Cerrillo J, Strasberg P and Schaller G 2018 {\em New J. Phys.\/}
  {\bf 20} 053063

\bibitem{SegalPRL2008}
Segal D 2008 {\em Phys. Rev. Lett.\/} {\bf 100} 105901

\bibitem{EspositoEtAlEPL2009}
Esposito M, Lindenberg K and Van~den Broeck C 2009 {\em Europhys. Lett.\/} {\bf
  85} 60010

\bibitem{SanchezButtikerPRB2011}
S{\'a}nchez R and B{\"u}ttiker M 2011 {\em Phys. Rev. B\/} {\bf 83} 085428

\bibitem{StrasbergEtAlPRL2013}
Strasberg P, Schaller G, Brandes T and Esposito M 2013 {\em Phys. Rev. Lett.\/}
  {\bf 110} 040601

\bibitem{SothmannEtAlNano2014}
Sothmann B, S{\'a}nchez R and Jordan A~N 2014 {\em Nanotechnology\/} {\bf 26}
  032001

\bibitem{BenentiEtAlPhysRep2017}
Benenti G, Casati G, Saito K and Whitney R~S 2017 {\em Phys. Rep.\/} {\bf 694}
  1--124

\bibitem{FeshchenkoEtAlPRB2014}
Feshchenko A~V, Koski J~V and Pekola J~P 2014 {\em Phys. Rev. B\/} {\bf 90}
  201407(R)

\bibitem{HartmannEtAlPRL2015}
Hartmann F, Pfeffer P, H{\"o}fling S, Kamp M and Worschech L 2015 {\em Phys.
  Rev. Lett.\/} {\bf 114} 146805

\bibitem{ThierschmannEtAllNano2015}
Thierschmann H, S{\'a}nchez R, Sothmann B, Arnold F, Heyn C, Hansen W, Buhmann
  H and Molenkamp L~W 2015 {\em Nat. Nanotechnol.\/} {\bf 10} 854--858

\bibitem{FilligerReimannPRL2007}
Filliger R and Reimann P 2007 {\em Phys. Rev. Lett.\/} {\bf 99}(23) 230602

\bibitem{ChianEtAlPRE2017}
Chiang K~H, Lee C~L, Lai P~Y and Chen Y~F 2017 {\em Phys. Rev. E\/} {\bf 96}(3)
  032123

\bibitem{RouletEtAlPRE2017}
Roulet A, Nimmrichter S, Arrazola J~M, Seah S and Scarani V 2017 {\em Phys.
  Rev. E\/} {\bf 95}(6) 062131

\bibitem{FogedbyImparatoEPL2018}
Fogedby H~C and Imparato A 2018 {\em Europhys. Lett.\/} {\bf 122} 10006

\bibitem{MarchegianiEtAlPRA2016}
Marchegiani G, Virtanen P, Giazotto F and Campisi M 2016 {\em Phys. Rev.
  Appl.\/} {\bf 6}(5) 054014

\bibitem{AlickiEtAlJPA2015}
Alicki R, Gelbwaser-Klimovsky D and Szczygielski K 2015 {\em J. Phys. A\/} {\bf
  49} 015002

\bibitem{AlickiEtAlAP2017}
Alicki R, Gelbwaser-Klimovsky D and Jenkins A 2017 {\em Ann. Phys.\/} {\bf 378}
  71--87

\bibitem{AlickiEntropy2016}
Alicki R 2016 {\em Entropy\/} {\bf 18} 210

\bibitem{SerraGarciaEtAlPRL2016}
Serra-Garcia M, Foehr A, Moler\'on M, Lydon J, Chong C and Daraio C 2016 {\em
  Phys. Rev. Lett.\/} {\bf 117}(1) 010602

\bibitem{Bang_etalNJP2018}
Bang J, Pan R, Hoang T~M, Ahn J, Jarzynski C, Quan H~T and Li T 2018 {\em New
  J. Phys.\/} {\bf 20} 103032

\bibitem{KochEtAlPRB2004}
Koch J, von Oppen F, Oreg Y and Sela E 2004 {\em Phys. Rev. B\/} {\bf 70}
  195107

\bibitem{GalperinEtAlPRB2009}
Galperin M, Saito K, Balatsky A~V and Nitzan A 2009 {\em Phys. Rev. B\/} {\bf
  80} 115427

\bibitem{RomanoEtAlPRB2010}
Romano G, Gagliardi A, Pecchia A and Di~Carlo A 2010 {\em Phys. Rev. B\/} {\bf
  81} 115438

\bibitem{SchulzeEtAlPRL2008}
Schulze G, Franke K~J, Gagliardi A, Romano G, Lin C, Rosa A, Niehaus T~A,
  Frauenheim T, Di~Carlo A, Pecchia A {\em et~al.\/} 2008 {\em Phys. Rev.
  Lett.\/} {\bf 100} 136801

\bibitem{ScorranoCarcaterraMSSSP2013}
Scorrano A and Carcaterra A 2013 {\em Mech. Syst. Signal Pr.\/} {\bf 39}
  489--514

\bibitem{IsacssonPRB2001}
Isacsson A 2001 {\em Phys. Rev. B\/} {\bf 64} 035326

\bibitem{GiaeverZellerPRL1968}
Giaever I and Zeller H 1968 {\em Phys. Rev. Lett.\/} {\bf 20} 1504

\bibitem{KulikShekhter1975}
Kulik I and Shekhter R 1975 {\em Zhur. Eksper. Teoret. Fiziki\/} {\bf 68}
  623--640

\bibitem{AverinLikharevJLTP1986}
Averin D and Likharev K 1986 {\em J. Low Temp. Phys.\/} {\bf 62} 345--373

\bibitem{LaiEtAlJoP2012}
Lai W, Cao Y and Ma Z 2012 {\em J. Phys. Condens. Matter\/} {\bf 24} 175301

\bibitem{LaiEtAlJoP2013}
Lai W, Xing Y and Ma Z 2013 {\em J. Phys. Condens. Matter\/} {\bf 25} 205304

\bibitem{FedoretsEtAlPRL2004}
Fedorets D, Gorelik L~Y, Shekhter R~I and Jonson M 2004 {\em Phys. Rev.
  Lett.\/} {\bf 92} 166801

\bibitem{BonetEtAlPRB2002}
Bonet E, Deshmukh M~M and Ralph D 2002 {\em Phys. Rev. B\/} {\bf 65} 045317

\bibitem{BagretsEtAlPRB2003}
Bagrets D and Nazarov Y~V 2003 {\em Phys. Rev. B\/} {\bf 67} 085316

\bibitem{HarbolaPRB2006}
Harbola U, Esposito M and Mukamel S 2006 {\em Phys. Rev. B\/} {\bf 74} 235309

\bibitem{KevorkianColeBook1996}
Kevorkian J and Cole J 1996 {\em Multiscale and Singular Perturbation Methods
  Applied Mathematical Sciences\/} (Springer-Verlag, New York)

\bibitem{NayfehBook2008}
Nayfeh A~H 2008 {\em Perturbation methods\/} (John Wiley \& Sons)

\bibitem{BenderOrszagBook2013}
Bender C~M and Orszag S~A 2013 {\em Advanced mathematical methods for
  scientists and engineers I: Asymptotic methods and perturbation theory\/}
  (Springer Science \& Business Media)

\bibitem{BlanterEtAlPRL2004}
Blanter Y~M, Usmani O and Nazarov Y~V 2004 {\em Phys. Rev. Lett.\/} {\bf 93}
  136802

\bibitem{RiskenBook1996}
Risken H 1996 {\em The Fokker-Planck Equation\/} (Springer)

\bibitem{EspositoLindenbergVandenBroeckEPL2009}
Esposito M, Lindenberg K and {Van den Broeck} C 2009 {\em Europhys. Lett.\/}
  {\bf 85} 60010

\bibitem{VanDenBroeckPCC2013}
Van~den Broeck C {\em et~al.\/} 2013 {\em Phys. Complex Colloids\/} {\bf 184}
  155--193

\bibitem{EspositoPRE2012}
Esposito M 2012 {\em Phys. Rev. E\/} {\bf 85} 041125

\bibitem{HerpichEtAlPRX2018}
Herpich T, Thingna J and Esposito M 2018 {\em Phys. Rev. X\/} {\bf 8} 031056

\bibitem{TonekaboniEtAlArXiv2018}
Tonekaboni B, Lovett B~W and Stace T~M 2018 {\em arXiv preprint
  arXiv:1809.04251\/}

\bibitem{CroyEisfeldEPL2012}
Croy A and Eisfeld A 2012 {\em Europhys. Lett.\/} {\bf 98} 68004

\bibitem{WisemanMilburnBook2010}
Wiseman H~M and Milburn G~J 2010 {\em Quantum Measurement and Control\/}
  (Cambridge: Cambridge University Press)

\end{thebibliography}

\appendix
\section{\label{secApp:Equivalence}Equivalence of Fokker-Planck equation and stochastic differential Equations}
In this section we will show that the FPE, Eq.~(\ref{eq:FullFPE}), and the stochastic differential Eqs.~(\ref{eq:StochasticEquations1})-(\ref{eq:StochasticEquations3}) describe the same system, i.e. both representations reproduce the same expectation values up to order $\mathcal O(dt)$. Taking an arbitrary differentiable scalar function $f$ one finds for the expectation value of $f(q,x,v)$ by employing the FPE, Eq.~(\ref{eq:FullFPE}), (assuming that boundary contributions vanish, i.e. $f(q,x,v)p(x,v,q,t) \to 0$ as $x\to\pm\infty$ or $v\to\pm\infty$)
\begin{equation}
\label{eqApp:ExpecationFPE}
\begin{aligned}
\frac{\partial\left<f(q,x,v)\right>}{\partial t} &= \sumint f(q,x,v)\left[-v\frac{\partial}{\partial x} + \frac{\partial}{\partial v}\left(\frac{k}{m}x+\frac{\gamma}{m}v - \frac{\alpha V}{m}q\right)+D\frac{\partial^2}{\partial v^2}\right]p(x,v,q,t) \\
&\quad +\sumint\sum\limits_{q'\nu} \left[f(q,x,v)R_{qq'}^\nu(x)p(x,v,q',t) - f(q,x,v)R_{q'q}^\nu(x)p(x,v,q,t)\right]\\
&=\sumint \left[\frac{\partial f(q,x,v)}{\partial x}v - \frac{\partial f(q,x,v)}{\partial v}\left(\frac{k}{m}x+\frac{\gamma}{m}v - \frac{\alpha V}{m}q\right)+D\frac{\partial^2 f(q,x,v)}{\partial v^2}\right]p \\
&\quad +\sumint\sum\limits_{q'\nu}   \left[f(q',x,v)-f(q,x,v)\right]R_{q'q}^\nu(x)p(x,v,q,t) \\
&=\left<\frac{\partial f}{\partial x}v  - \frac{\partial f}{\partial v}\left(\frac{k}{m}x+\frac{\gamma}{m}v - \frac{\alpha V}{m}q\right)  + \frac{\partial^2 f}{\partial v^2}D\right>  +\left<\sum\limits_{\substack{q'\nu}}\left[f(q')-f(q)\right]R_{q'q}^\nu\right>, 
\end{aligned}
\end{equation}
where $\sumint \equiv \int dx \int dv \sum_q$. Our aim is to show that Eq.~(\ref{eqApp:ExpecationFPE}) is also obtained by the use of the stochastic differential Eqs.~(\ref{eq:StochasticEquations1})-(\ref{eq:StochasticEquations3}). In terms of the stochastic process defined by Eqs.~(\ref{eq:StochasticEquations1})-(\ref{eq:StochasticEquations3}) we can write for the expectation value of the increment of the arbitrary scalar function $f$ the following (using It\^o's Lemma):
\begin{equation}
\label{eqApp:firstStochastic}
\begin{aligned}
\mathbb E\left[f(q+dq,x+dx,v+dv)-f(q,x,v)\right] =& \mathbb E\left[\frac{\partial f}{\partial x}v\right] dt - \mathbb E\left[\frac{\partial f}{\partial v}\left(\frac{k}{m}x+\frac{\gamma}{m}v - \frac{\alpha V}{m}q\right)\right] dt \\
&+ \mathbb E\left[\frac{\partial^2 f}{\partial v^2}D\right]dt  +\mathbb E\left[\sum\limits_{k=1}^\infty \frac{1}{k!}f^{(k)}\left(dq\right)^k\right] + \mathcal O(dt^2),
\end{aligned}
\end{equation}
where we have used the statistical properties of the Wiener increment, i.e., $\mathbb E\left[dB(t)\right] = 0$ and $\mathbb E\left[(dB(t))^2\right] = dt$. Note that mixed terms of $dB(t)$ and $dN_{q'q}^\nu(x,t)$ exceed the leading order of $dt$ because $\mathbb E\left[dN_{q'q}^\nu(x, t)\right]\propto dt$. Here, $\mathbb E\left[\bullet\right]$ denotes averages of the stochastic process. We now evaluate the sum in Eq.~(\ref{eqApp:firstStochastic}): Since all powers of the Poisson increment are of order $dt$, we have to evaluate the sum exactly. Since $\left[dN_{q'q}^\nu(x,t)\right]^k = dN_{q'q}^\nu(x,t)$ for all $k\in \mathbb N$ we can rewrite the expectation value as follows (omitting any dependencies on $x$ and $v$):
\begin{equation}
\begin{aligned}
\mathbb E\left[\sum\limits_{k=1}^\infty \frac{f^{(k)}(q)}{k!}\left(dq\right)^k\right]&=\mathbb E\left[\sum\limits_{k=1}^\infty \frac{f^{(k)}(q)}{k!}\sum\limits_{\substack{q'\nu}} \left(q'-q\right)^k dN_{q'q}^\nu\right] =\mathbb E\left[\sum\limits_{\substack{q'\nu}}\left\{f(q')-f(q)\right\}dN_{q'q}^\nu\right]\\&=\mathbb E\left[\sum\limits_{\substack{q'\nu}}\left\{f(q')-f(q)\right\}R_{q'q}^\nu\right]dt, 
\end{aligned}
\end{equation}
where the last equality follows from a general identity of point/Poisson processes (see, e.g., Eq.~(B.54) in \cite{WisemanMilburnBook2010}). Hence, up to order $\mathcal O(dt)$ we write
\begin{equation}
\label{eq:StochasticAverageApp}
\begin{aligned}
\frac{\partial}{\partial t}\mathbb E\left[f(q,x,v)\right] =&\mathbb E\left[\frac{\partial f}{\partial x}v\right]  - \mathbb E\left[\frac{\partial f}{\partial v}\left(\frac{k}{m}x+\frac{\gamma}{m}v - \frac{\alpha V}{m}q\right)\right] + \mathbb E\left[\frac{\partial^2 f}{\partial v^2}D\right] \\
&+\mathbb E\left[\sum\limits_{\substack{q'\nu}}\left[f(q')-f(q)\right]R_{q'q}^\nu\right].
\end{aligned}
\end{equation}
Since Eq.~(\ref{eqApp:ExpecationFPE}) is equivalent to Eq.~(\ref{eq:StochasticAverageApp}) we can conclude that, for the same initial conditions, expectation values of an arbitrary function $f$ with respect to the probability density and with respect to realizations of the stochastic process evolve equally. Hence, the FPE, Eq.~(\ref{eq:FullFPE}), and the stochstic differential Eqs.~(\ref{eq:StochasticEquations1})-(\ref{eq:StochasticEquations3}) describe the same process.

\section{\label{secApp:MultipleScale}Multiple scale perturbation theory}
In this section we derive Eq.~(\ref{eq:FokkerPlanckHO}) from the full FPE, Eq.~(\ref{eq:FullFPE}). The idea of MS perturbation theory is to impose a time scale separation of the frequent electron tunnelling events and the slow evolution of the oscillator and, furthermore, to demand that those terms of the approximated solution that grow with time, vanish. By imposing the latter condition we ensure that the MS solution of the full probability density will be valid on the long time scale.

First we will state some useful properties of the matrix $R(x)=\sum_\nu R^\nu(x)$  [see Eq.~(\ref{eq:Rate})], which we will use throughout the derivation. Since $R(x)$ is a $2\times 2$ rate matrix it has two eigenvalues: $0$ and $\chi<0$. Accordingly, there are two (right) eigenvectors, $\boldsymbol \pi(x)$ and $\boldsymbol \chi$, for which $R(x)\boldsymbol\pi(x) = 0$ and $R(x)\boldsymbol \chi = \chi\boldsymbol\chi$ holds, respectively. Here, 
\begin{equation}
\label{eqApp:RightPiEV}
\boldsymbol \pi(x) = \left(\begin{array}{c} 
\pi_0(x) \\ 
\pi_1(x) \end{array}\right)
\end{equation}
is the (instantaneous) stationary solution of $R(x)$. Note that since $R = R(x)$, the stationary state $\boldsymbol \pi(x)$ is also a function of the position $x$ and we impose the normalization condition $\pi_0(x) + \pi_1(x) = 1$. Furthermore, 
\begin{equation}
\label{eqApp:RightChiEV}
\boldsymbol \chi = \left(\begin{array}{c}1 \\-1 \end{array}\right)
\end{equation}
and
\begin{equation}
\chi = -2 \Gamma\cosh\left(\frac{x}{\lambda}\right).
\end{equation}
There are additionally left eigenvectors 
\begin{equation}
\label{eqApp:LeftPiEV}
\boldsymbol \pi^\dagger = \left(1,1\right)
\end{equation}
 and 
\begin{equation}
\label{eqApp:LeftChiEV}
\boldsymbol \chi^\dagger(x) = (\pi_1(x),-\pi_0(x)).
\end{equation}
satisfying $\boldsymbol \pi^\dagger R(x) = 0$ and $\boldsymbol \chi^\dagger(x) R(x) = \chi\boldsymbol\chi^\dagger(x)$.

We start the derivation of Eq.~(\ref{eq:FokkerPlanckHO}) by considering Eq.~(\ref{eq:FullFPE}) in matrix representation, i.e.
\begin{equation}
\label{eq:FullMasterEquation}
\begin{aligned}
\frac{\partial \pb}{\partial t}=\left[ -v \frac{\partial}{\partial x}+\frac{\partial}{\partial v}\left(\frac{k}{m} x+ \frac{\gamma}{m} v\right)\right]\pb + \frac{\alpha V}{m} \left(\begin{array}{cc}
0 & 0\\
0 & -1
\end{array}\right) \frac{\partial \pb}{\partial v}+ D\frac{\partial^2 \pb}{\partial v^2} +R(x)\pb.
\end{aligned}
\end{equation}
where $\pb=(p_0(x,v,t),p_1(x,v,t))^\intercal$. Introducing the differential operator $\mathcal L$
\begin{equation}
\label{eq:DefinitionDiffOp}
\begin{aligned}
\mathcal L = \left[ -v \frac{\partial}{\partial x}+\frac{\partial}{\partial v}\left(\frac{k}{m} x+ \frac{\gamma}{m} v\right)\right]+ \frac{\alpha V}{m} \left(\begin{array}{cc}
0 & 0\\
0 & -1
\end{array}\right) \frac{\partial }{\partial v}+ D\frac{\partial^2 }{\partial v^2},
\end{aligned}
\end{equation}
Eq.~(\ref{eq:FullMasterEquation}) takes the compact form
\begin{equation}
\label{eq:FPEinL}
\frac{\partial \pb}{\partial t} (x,v,t)= \mathcal L \pb(x,v,t) + R\pb(x,v,t).
\end{equation}
Since the fast time scale describes the dynamics of the two state system, we treat the $\mathcal L$-part as perturbation and introduce the bookkeeping parameter $\epsilon \ll 1$ such that we can rewrite Eq.~(\ref{eq:FPEinL}) as
\begin{equation}
\label{eq:InitialPerturbedEquation}
\frac{\partial \pb}{\partial t}= \epsilon \mathcal L\pb + R\pb.
\end{equation}
The idea of MS perturbation theory is now to introduce two time scales, a fast one ($\tilde t$) and a slow one $\tau = \epsilon \tilde t$ such that the probability density is a function of both times scales, i.e. $\tilde \pb(x,v,\tilde t,\tau) = \pb(x,v,t) $. The temporal derivative transforms to a sum provided by the chain rule:
\begin{equation}
\frac{\partial}{\partial t} = \frac{\partial}{\partial \tilde t} + \epsilon\frac{\partial}{\partial \tau}.
\end{equation}
Then, Eq.~(\ref{eq:InitialPerturbedEquation}) is given by
\begin{equation}
\frac{\partial \tilde \pb}{\partial \tilde t}+\epsilon \frac{\partial \tilde \pb}{\partial \tau} = \epsilon\mathcal L\tilde  \pb + R\tilde \pb.
\end{equation}
From this point on, we will refer to $\tilde t$ as $t$ and to $\tilde \pb(x,v,t,\tau)$ as $\pb(t,\tau)$. 
Assuming that we can express $\pb$ as a series of orders of $\epsilon$, 
\begin{equation}
\pb(t,\tau) = \pbn 0(t,\tau) + \epsilon \pbn 1(t,\tau) + \epsilon^2 \pbn 2(t,\tau) + \mathcal O(\epsilon^3),
\end{equation}
we find a hierarchy of equations for the different orders of $\epsilon$. The goal of the MS perturbation theory is now to find an approximate solution, such that, after setting $\epsilon$ to 1, it holds 
\begin{equation}
\pb(t,\tau) \approx \pbn 0(t,\tau) + \pbn 1(t,\tau).
\end{equation}

We start with the governing equation for $\mathcal O(\epsilon^0)$:
\begin{equation}
\label{eq:ZeroOrderODE}
\frac{\partial \pbn 0}{\partial t}(t,\tau) = R \pbn 0(t,\tau).
\end{equation}
The simplest solution of the ordinary differential Eq.~(\ref{eq:ZeroOrderODE}) is given by assuming that the left hand side of Eq.~(\ref{eq:ZeroOrderODE}) is equal to $0$, i.e. assuming that the probability density at zeroth order is independent of the fast 
time scale $t$: $\pbn 0(t,\tau) = \pbn 0(\tau)$. This means $\pbn 0(\tau)$ must be the eigenvector of $R$ with eivenvalue $0$, i.e. 
\begin{equation}
\label{eq:ZerothOrderSolution}
\pbn 0 (\tau) = \boldsymbol \pi(x) \pn 0 (\tau).
\end{equation}
Here, $\pn 0 (\tau)$ is a scalar function which represents the probability density of the oscillator alone, i.e. tracing out the charge state $q$ of $\pbn 0(x,v,\tau)$ results in 
\begin{equation}
\pi_0(x)\pn 0 (x,v,\tau) + \pi_1(x)\pn 0(x,v,\tau) = \pn 0(x,v,\tau), 
\end{equation}
which is the probability density to find the oscillator at position $x$ with velocity $v$ at time $\tau$ (at zeroth order). The specific form of $\pn 0 (\tau)$ will be determined by the first order of the perturbation hierarchy.  Note that, if we stop the perturbation theory here, Eq.~(\ref{eq:ZerothOrderSolution}) implies an infinite time scale separation, which is equivalent to an adiabatic approximation. 

The equation of motion of $\pb(t,\tau)$ at $\mathcal O(\epsilon)$ is given by
\begin{equation}
\label{eq:FirstOrderEquationOfMotion}
\frac{\partial \pbn 1}{\partial t}(t,\tau) = -\frac{\partial \pbn 0(\tau)}{\partial \tau} + \mathcal L \pbn 0(\tau) + R\pbn 1(t,\tau).
\end{equation}
Since $\pbn 0 (x,v,\tau)$ is independent of $t$ [see Eq.~(\ref{eq:ZerothOrderSolution})], the solution of the latter equation is formally given by
\begin{equation}
\label{eq:FirstOrderSolution}
\begin{aligned}
\pbn 1(t,\tau) = e^{R t} \pbnt 1(\tau) + e^{Rt}\int\limits_0^{t}e^{-R s} ds~\left[-\frac{\partial \pbn 0(\tau)}{\partial \tau}+\mathcal L \pbn 0(\tau) \right], 
\end{aligned}
\end{equation}
where $\pbnt 1(\tau)$ is a probability vector which does not depend on the fast time scale $t$ but is 
unspecified at this moment. 

Next we look at the integral of Eq.~(\ref{eq:FirstOrderSolution}): We can expand the exponential of the rate matrix by use of the eigenvalues and eigenvectors of $R$ [see Eqs.~(\ref{eqApp:RightPiEV})-(\ref{eqApp:LeftChiEV})], i.e.
\begin{equation}
\int\limits_0^t e^{-Rs}ds = \int\limits_0^t \boldsymbol \pi(x) \boldsymbol \pi^\dagger +e^{-\chi s} \boldsymbol \chi\boldsymbol \chi^\dagger ds
\end{equation}
Evaluating the integral gives
\begin{equation}
\label{eq:SolutionOfIntegral}
\int \limits_0^t e^{-Rs}ds = \boldsymbol\pi(x)\boldsymbol\pi^\dagger t - \frac{e^{-\chi t}-1}{\chi}\boldsymbol \chi \boldsymbol\chi^\dagger.
\end{equation}
Inserting Eq.~(\ref{eq:SolutionOfIntegral}) into Eq.~(\ref{eq:FirstOrderSolution}) we find that there are terms in the 
solution which grow linearly with $t$ for long times, i.e.
\begin{equation}
e^{Rt}\boldsymbol\pi(x)\boldsymbol\pi^\dagger t \left[-\frac{\partial \pbn 0}{\partial \tau} + \mathcal L \pbn 0\right].
\end{equation}
Those terms, which will subsequently be referred to as secular terms, prohibit a steady state solution of the perturbation hierarchy. We therefore demand the secular terms to vanish such that we find a stable solution. 

At the first order perturbation the latter condition is satisfied if 
\begin{equation}
\label{eq:ConditionFirstOrder}
\boldsymbol\pi^\dagger \left[-\frac{\partial \pbn 0}{\partial \tau}+\mathcal L \pbn 0\right] = 0. 
\end{equation}
Inserting $\pbn 0$ [see Eq.~(\ref{eq:ZerothOrderSolution})] yields
\begin{equation}
\boldsymbol \pi^\dagger \left[-\boldsymbol\pi(x) \frac{\partial \pn 0 }{\partial \tau}+\mathcal L \boldsymbol\pi(x)\pn 0 \right]=0. 
\end{equation}
It holds that $\boldsymbol\pi^\dagger \boldsymbol\pi(x) = 1$ [see Eqs.(\ref{eqApp:RightPiEV}) and (\ref{eqApp:LeftPiEV})]. Therefore
\begin{equation}
-\frac{\partial \pn 0 }{\partial \tau}+\boldsymbol\pi^\dagger\mathcal L \boldsymbol\pi(x)\pn 0 =0. 
\end{equation}
Evaluating the action of the differential operator $\mathcal L$ on $\boldsymbol\pi(x)$ results in 
\begin{equation}
\label{eq:piLpi}
\begin{aligned}
\boldsymbol\pi^\dagger\mathcal L \boldsymbol\pi(x) =-v\frac{\partial}{\partial x}+\frac{\partial}{\partial v}\left(\frac{k}{m}x+\frac{\gamma}{m} v\right) -\frac{\alpha V}{m}q_{\text{eq}}(x)\frac{\partial}{\partial v} + D\frac{\partial^2}{\partial v^2},
\end{aligned}
\end{equation}
where $q_{\text{eq}}(x) = \pi_1(x)$. Furthermore, we have used that [see Eqs.~(\ref{eqApp:RightPiEV}) and (\ref{eqApp:LeftPiEV})]
\begin{equation}
\begin{aligned}
\boldsymbol\pi^\dagger \frac{\partial}{\partial x}\boldsymbol\pi(x) = \frac{\partial}{\partial x} \boldsymbol\pi^\dagger \boldsymbol\pi(x) = \frac{\partial}{\partial x} 1 = 0.
\end{aligned}
\end{equation}

The condition for secular terms to vanish in the first order perturbation [see Eq.~(\ref{eq:ConditionFirstOrder})] is now given by
\begin{equation}
\label{eq:FPE0order}
\begin{aligned}
\frac{\partial \pn 0}{\partial \tau}=-v\frac{\partial}{\partial x}\pn 0 + \frac{\partial}{\partial v}\left[\frac{k}{m} x+ \frac{\gamma}{m} v - \frac{\alpha V}{m}q_{\text{eq}}(x)\right]\pn 0 + D\frac{\partial^2}{\partial v^2}\pn 0.
\end{aligned}
\end{equation}
The latter equation is a FPE for $\pn 0 (x,v,\tau)$ describing the underdamped evolution in an effective potential $U_{\text{eff}}(x)$ with $\partial_x U_{\text{eff}} = kx-\alpha V q_{\text{eq}}(x)$. This corresponds to our ansatz for the $0$th order, where we have assumed that the QD is in its instantaneous equilibrium state at all times. The harmonic potential is therefore altered and the effective FPE describing the oscillator is simple diffusion within the effective potential.

We now return to the perturbation of $\mathcal O(\epsilon)$ [Eq.~{\ref{eq:FirstOrderSolution}]: After removing the secular terms, the first order solution is given by
\begin{equation}
\label{eq:FirstOrderStep}
\pbn 1 = e^{Rt}\pbnt 1 + e^{Rt}\boldsymbol \chi\boldsymbol\chi^\dagger\frac{1-e^{-\chi t}}{\chi}\left[-\frac{\partial \pbn 0}{\partial \tau}+\mathcal L\pbn 0\right].
\end{equation}
The exponential of the rate matrix $R$ can again be expressed in terms of the eigenvectors,
\begin{equation}
\label{eq:ExpansionRate}
e^{Rt}=\boldsymbol\pi(x)\boldsymbol\pi^\dagger+e^{\chi t}\boldsymbol\chi \boldsymbol\chi^\dagger. 
\end{equation}
Inserting Eq.~(\ref{eq:ExpansionRate}) into Eq.~(\ref{eq:FirstOrderStep}) results in
\begin{equation}
\label{eq:FirstOrderStep2}
\begin{aligned}
\pbn 1 =& \boldsymbol\pi(x)\boldsymbol\pi^\dagger\pbnt 1 +e^{\chi t}\boldsymbol\chi \boldsymbol\chi^\dagger\pbnt 1 + \boldsymbol\pi(x)\boldsymbol\pi^\dagger\boldsymbol \chi\boldsymbol\chi^\dagger\frac{1-e^{-\chi t}}{\chi}\left[-\frac{\partial \pbn 0}{\partial \tau}+\mathcal L\pbn 0\right]\\
&+\boldsymbol \chi\boldsymbol\chi^\dagger\frac{e^{\chi t}-1}{\chi} \left[-\frac{\partial \pbn 0}{\partial \tau}+\mathcal L\pbn 0\right],
\end{aligned}
\end{equation}
where we have used that $\boldsymbol \chi^\dagger \boldsymbol \chi =1$ [see Eq.(\ref{eqApp:LeftChiEV}) and (\ref{eqApp:RightChiEV})]. In the long-time limit terms proportional to $e^{\chi t}$ will approach zero, since $\chi<0$. Using the latter as well as the fact that $\boldsymbol\pi^\dagger \boldsymbol\chi = 0$, Eq.~(\ref{eq:FirstOrderStep2}) simplifies to
\begin{equation}
\pbn 1 = \boldsymbol\pi(x)\pn 1 -\boldsymbol \chi\boldsymbol\chi^\dagger\frac{1}{\chi} \left[-\frac{\partial \pbn 0}{\partial \tau}+\mathcal L\pbn 0\right],
\end{equation} 
where $\pn 1 \equiv \boldsymbol\pi^\dagger\pbnt 1$. Substituting now $\pbn 0(\tau)$ by $\boldsymbol \pi(x) \pn 0(\tau)$ [see Eq.~(\ref{eq:ZerothOrderSolution})] and the differential operator $\mathcal L$ by its definition [see Eq.~(\ref{eq:DefinitionDiffOp})] we can approximate the first order solution [Eq.~(\ref{eq:FirstOrderStep})] by
\begin{equation}
\label{eq:firstOrderSolution}
\begin{aligned}
\pbn 1  =&\boldsymbol \pi(x)\pn 1-\boldsymbol \chi \boldsymbol \chi^\dagger \frac{1}{\chi}\left\{-v \frac{\partial \boldsymbol\pi(x)}{\partial x}+\frac{\alpha V}{m}\left(\begin{array}{c}
0\\
-\pi_{1}(x)
\end{array}\right)\frac{\partial}{\partial v}\right\}\pn 0,
\end{aligned}
\end{equation}
using $\boldsymbol\chi^\dagger \boldsymbol\pi(x)=0$ [see Eqs.~(\ref{eqApp:RightPiEV}) and (\ref{eqApp:LeftChiEV})]. 

We now define a new vector
\begin{equation}
\boldsymbol\kappa \equiv \left\{-v \frac{\partial \boldsymbol\pi(x)}{\partial x}+\frac{\alpha V}{m}\left(\begin{array}{c}
0\\
-\pi_{1}(x)
\end{array}\right)\frac{\partial}{\partial v}\right\}\pn 0
\end{equation}
such that Eq.~(\ref{eq:firstOrderSolution}) becomes
\begin{equation}
\pbn 1 = \boldsymbol \pi(x)\pn 1-\frac{1}{\chi}\boldsymbol\chi \boldsymbol\chi^\dagger \boldsymbol\kappa
\end{equation}
Note that $\boldsymbol\chi^\dagger \boldsymbol\kappa\neq 0$ in general.

Similar to the procedure for the zeroth order perturbation we now look at the second order of $\pb$ and demand secular terms to vanish. By this condition we will find a differential equation for $\pn 1$, which describes the effective evolution of the oscillator at a first order perturbation level without taking the electronic degrees of freedom specifically into account. The governing equation at $\mathcal O(\epsilon^2)$ is similar to Eq.~(\ref{eq:FirstOrderEquationOfMotion}) and reads
\begin{equation}
\frac{\partial \pbn 2}{\partial t} = -\frac{\partial \pbn 1}{\partial \tau} + \mathcal L \pbn 1 + R\pbn 2.
\end{equation}
Again, the general solution can be written as
\begin{equation}
\pbn 2 = e^{Rt}\pbnt 2 + e^{Rt}\int\limits_0^t e^{-Rs}ds \left[-\frac{\partial \pbn 1}{\partial \tau}+\mathcal L\pbn 1\right].
\end{equation}
With the same calculation as above we find that in order for the secular terms in the second order to vanish, the following condition must hold:
\begin{equation}
\label{eq:SecularitiesFirstOrder}
\begin{aligned}
\boldsymbol\pi^\dagger \left[-\frac{\partial \pbn 1}{\partial \tau}+\mathcal L \pbn 1\right] &=0 \\
\Leftrightarrow -\frac{\partial \pn 1}{\partial \tau}+ \boldsymbol \pi^\dagger \mathcal L \boldsymbol \pi(x)\pn 1 - \frac{1}{\chi}\boldsymbol\pi^\dagger \mathcal L\boldsymbol \chi \boldsymbol \chi^\dagger \boldsymbol \kappa &=0.
\end{aligned}
\end{equation}
The term $\boldsymbol \pi^\dagger\mathcal L\boldsymbol \pi(x)$ appears again and is given by Eq.~(\ref{eq:piLpi}).
We now evaluate the third term on the left hand side of Eq.~(\ref{eq:SecularitiesFirstOrder}): First we note that $\boldsymbol \pi^\dagger \boldsymbol \chi = 0$ [see Eqs.~(\ref{eqApp:RightChiEV}) and (\ref{eqApp:LeftPiEV})]. Therefore, we only have to evaluate the action of $\mathcal L$ on $\boldsymbol \chi$ in Eq.~(\ref{eq:SecularitiesFirstOrder}), which is 
\begin{equation}
\mathcal L\boldsymbol\chi= \frac{\alpha V}{m} \left(\begin{array}{c}
0 \\
1
\end{array}\right) \frac{\partial}{\partial v},
\end{equation}
because $\boldsymbol \chi$ is a constant vector [see Eq.~(\ref{eqApp:RightChiEV})]. 
It then holds [see Eq.~(\ref{eqApp:LeftPiEV}) and (\ref{eqApp:LeftChiEV})]
\begin{equation}
\begin{aligned}
\frac{1}{\chi}\boldsymbol \pi^\dagger \mathcal L\boldsymbol\chi \boldsymbol \chi^\dagger \boldsymbol \kappa &=\frac{\alpha V}{\chi m}\boldsymbol \pi^\dagger\left(\begin{array}{c}
0 \\
1
\end{array}\right)\frac{\partial}{\partial v} \left(\boldsymbol \chi^\dagger \boldsymbol\kappa\right)\\
& = \frac{\alpha V}{\chi m}\boldsymbol \chi^\dagger \frac{\partial }{\partial v}\boldsymbol \kappa.
\end{aligned}
\end{equation}

Lastly, we look at the derivative of $\boldsymbol \kappa$ with respect to $v$, that is
\begin{equation}
\frac{\partial }{\partial v}\boldsymbol \kappa=-\left(\frac{\partial }{\partial x}\boldsymbol \pi(x)\right)\frac{\partial}{\partial v}(v\pn 0) + \frac{\alpha V}{m}\left(\begin{array}{c}
0\\
-\pi_1
\end{array}\right)\frac{\partial^2\pn 0}{\partial v^2}.
\end{equation}
Using $\pi_1 = 1-\pi_0 = q_{\text{eq}}$,} one can show that 
\begin{equation}
\pi_0\frac{\partial \pi_1}{\partial x}-\pi_1\frac{\partial \pi_0}{\partial x} = \frac{\partial q_{\text{eq}}}{\partial x}.
\end{equation}
Furthermore it holds that $\pi_0\pi_1 = q_{\text{eq}}(x)-q_{\text{eq}}^2(x)$. With the latter two simplifications we can rewrite [see Eq.~(\ref{eqApp:LeftChiEV})]:
\begin{equation}
\boldsymbol\chi^\dagger \left(\frac{\partial }{\partial v}\boldsymbol\kappa\right) = \frac{\partial q_{\text{eq}}}{\partial x}\frac{\partial}{\partial v}\left(v \pn 0\right) + \frac{\alpha V}{m}q_{\text{eq}}\left(1-q_{\text{eq}}\right)\frac{\partial^2 \pn 0}{\partial v^2}.
\end{equation}
Putting everything together, the condition for the secular terms of the solution at second order in the perturbation [see Eq.~(\ref{eq:SecularitiesFirstOrder})] is given by
\begin{equation}
\label{eq:FirstOrderFPE}
\begin{aligned}
\frac{\partial \pn 1}{\partial \tau} =&\mathcal L_0\pn 1 - \frac{\alpha V}{\chi m}\left[\frac{\partial q_{\text{eq}}}{\partial x}\frac{\partial (v \pn 0)}{\partial v}+\frac{\alpha V q_{\text{eq}}}{ m}\left(1-q_{\text{eq}}\right)\frac{\partial^2\pn 0}{\partial v^2}\right]
\end{aligned}
\end{equation}
where
\begin{equation}
\mathcal L_0 \equiv -v \frac{\partial}{\partial x}+\frac{\partial}{\partial v}\left(\frac{k}{m}x + \frac{\gamma}{m} v - \frac{\alpha V}{m}q_{\text{eq}}\right) + D\frac{\partial^2}{\partial v^2}.
\end{equation}

Putting zeroth and first order together [see Eqs.~(\ref{eq:FPE0order}) and (\ref{eq:FirstOrderFPE})], i.e., $\tilde p(x,v,\tau) = \pn 0(x,v,\tau) + \epsilon \pn 1(x,v,\tau)$ and setting $\epsilon$ to $1$ ($\tau = \epsilon t \to t$), we find that the full probability density $p(x,v,q,t)$ can be approximated by 
\begin{equation}
p(x,v,q,t) \approx \pi_q(x)\tilde p(x,v,t),
\end{equation}
where $\tilde p(x,v,t)$ is the probability density of solely the oscillator, obtained by tracing out the fast electronic degrees of freedom. The dynamics of $p(x,v,t)$ is governed by a FPE:
\begin{equation}
\label{eq:FinalFPE}
\begin{aligned}
\frac{\partial \tilde p}{\partial t} =& \left[-v \frac{\partial}{\partial x} + \frac{\partial}{\partial v}\left(\frac{k}{m}x + \frac{\tilde \gamma(x)}{m}v\right) - \frac{\alpha V q_{\text{eq}}(x)}{m} \frac{\partial}{\partial v}\right]\tilde p + \tilde D(x) \frac{\partial^2}{\partial v^2}\tilde p,
\end{aligned}
\end{equation}
where the effective friction and diffusion coefficients are now position dependent and are given by
\begin{equation}
\begin{aligned}
\tilde \gamma(x) &= \gamma - \frac{\alpha V}{ \chi}\frac{\partial q_{\text{eq}}(x)}{\partial x},\\
\tilde D(x) &=  D-\frac{\alpha^2 V^2 q_{\text{eq}}(x)}{m^2 \chi(x) }\left(1-q_{\text{eq}}(x)\right).
\end{aligned}
\end{equation}
The final Eq.~(\ref{eq:FinalFPE}) is equivalent to Eq.~(\ref{eq:FokkerPlanckHO}) of the main text.

\section{\label{secApp:PolarCoordinates} Transformation to energy space}

In this section we derive the transformed FPE, Eq.~(\ref{eq:FPEEnergy}), from Eq.~(\ref{eq:FokkerPlanckHO}). We first rewrite Eq.~(\ref{eq:FokkerPlanckHO}) as follows:
\begin{equation}
\label{eqApp:FPEHO}
\frac{\partial \tilde p }{\partial t}  = -v \frac{\partial\tilde p}{\partial x} + \frac{\tilde \gamma(x)}{m}\tilde p+ \left(\frac{k}{m}x + \frac{\tilde \gamma(x)}{m}v - \frac{\alpha V }{m}q_{\text{eq}}(x) \right)\frac{\partial \tilde p }{\partial v}+ \tilde D(x) \frac{\partial^2 \tilde p}{\partial v^2}.
\end{equation}
Using the transformation of Eq.~(\ref{eq:Transformation}) as well as the assumption $\hat p(\mathcal E,\theta,t) \approx \hat p(\mathcal E,t)$, we find that derivatives with respect to $x$ and $v$ transform as follows:
\begin{equation}
\begin{aligned}
\frac{\partial \tilde p}{\partial x} &\to \sqrt{2\mathcal Ek}\sin\theta\frac{\partial \hat p}{\partial \mathcal E},\\
\frac{\partial \tilde p}{\partial v} &\to \sqrt{2\mathcal Em}\cos\theta \frac{\partial \hat p}{\partial \mathcal E},\\
\frac{\partial^2 \tilde p}{\partial v^2} &\to m\frac{\partial \hat p}{\partial \mathcal E}+2\mathcal Em\cos^2\theta\frac{\partial^2\hat p}{\partial \mathcal E^2}.
\end{aligned}
\end{equation}
In energy space, Eq.~(\ref{eqApp:FPEHO}) then takes the form
\begin{equation}
\label{eqApp:BeforeAveraging}
\begin{aligned}
\frac{\partial}{\partial t}\hat p(\mathcal E,t) =& \left[\frac{\tilde \gamma(x)}{m} + \left(\frac{\tilde \gamma(x)}{m}2\mathcal E\cos^2\theta -\alpha V q_\text{eq}(x)\sqrt{\frac{2\mathcal E}{m}}\cos\theta+\tilde D(x)m\right)\frac{\partial}{\partial \mathcal E}\right.\\
&+\left.\tilde D(x)2\mathcal E m\cos^2\theta \frac{\partial^2}{\partial\mathcal E^2}\right]\hat p(\mathcal E,t),
\end{aligned}
\end{equation}
where $x=\sqrt{2 \mathcal E/k}\sin \theta$. Upon averaging over $\theta$, the term proportional to $\cos\theta$ will vanish: As $q_\text{eq}(x)$ is an analytic function of $x$ it can be Taylor expanded in a power series and the individual contributions of all terms vanish due to $\int_0^{2\pi} \sin^n(\theta) \cos(\theta) d\theta = 0$ for all $n\in\mathbb N$. By inspection of Eq.~(\ref{eq:TransformedGammaE}) and partial integration one can show that the following relations hold:
\begin{equation}
\begin{aligned}
\frac{1}{2\pi}\int\limits_0^{2\pi}d\theta \tilde \gamma(x) &= 2 \hat\gamma(\mathcal E) + 2\mathcal E \frac{\partial \hat \gamma(\mathcal E)}{\partial \mathcal E},\\
\frac{1}{2\pi}\int\limits_0^{2\pi}d\theta \tilde D(x) &= 2 \hat D(\mathcal E) + 2\mathcal E \frac{\partial \hat D(\mathcal E)}{\partial \mathcal E}.\\
\end{aligned}
\end{equation}
Then, averaging Eq.~(\ref{eqApp:BeforeAveraging}) results in Eq.~(\ref{eq:FPEEnergy}).

\section{\label{secApp:ComputationalMethods} Computational methods}
For all numerical investigations we set $\beta^{\text{osc}} = \beta^\nu =\beta \equiv 1$ as well as $\Gamma \equiv 1$ and $\lambda \equiv 1$. The other parameters used in this work are given in units of the latter three: $\alpha \lambda = 0.06$, $m \lambda^2 \Gamma^2 \beta = 12.0$, $k\lambda^2 \beta=5.0$ and $\gamma \lambda^2\Gamma \beta = 0.2$. 

Since the probability space of the coupled system of harmonic oscillator and QD is very large, we assume that 
the system is ergodic, such that we can sample the steady state probability density of the system by a single long trajectory. Additionally this means that an ensemble average of an arbitrary quantity $A$ in the steady state is calculated by 
\begin{equation}
\left<A\right> = \frac{1}{T}\int\limits_0^T A(t),
\end{equation}
which is exact for ergodic systems in the limit of $T\to\infty$. 
We simulate the trajectories after a relaxation time of $\Gamma t=1000$ until $\Gamma T = 5000000$, where we have also checked that further relaxation time or simulation time does not change the probability density or averaged quantities. Note that we have also investigated different initial conditions and have not seen any dependency of the outcome on the initial conditions (after the relaxation time). Finally we note that the time step used in the simulations is $\Gamma \Delta t = 0.0001$.

\section{\label{subsec:Hopf bifurcation}Hopf bifurcation of the mean-field model}
In order to determine the critical value $\bar V_{\text{cr}}$, for which the electron shuttle bifurcates from a stable fixed point into a stable limit cycle, we perform a linear stability analysis around the fixed point of the MF Eqs.~(\ref{eq:MeanField1})-(\ref{eq:MeanField3}):
\begin{equation}
\begin{aligned}
\dot{\bar x} &= \bar{v}, \\
\dot{\bar v} &= -\frac{k}{m}\bar x - \frac{\gamma}{m} \bar v + \frac{\alpha V}{m} \bar p_1,\\
\dot{\bar p}_1 &= \sum\limits_\nu R^\nu_{10}(\bar x) \left(1-\bar p_{1}\right) - R_{01}^\nu (\bar x)\bar p_{1},
\end{aligned}
\end{equation}
where we have eliminated one equation compared to Eqs.~(\ref{eq:MeanField1})-(\ref{eq:MeanField3}) by use of probability conservation, i.e. $\bar p_{0} + \bar p_{1} = 1$. 

\begin{figure}[h]
\begin{center}
\includegraphics[width=0.7\columnwidth]{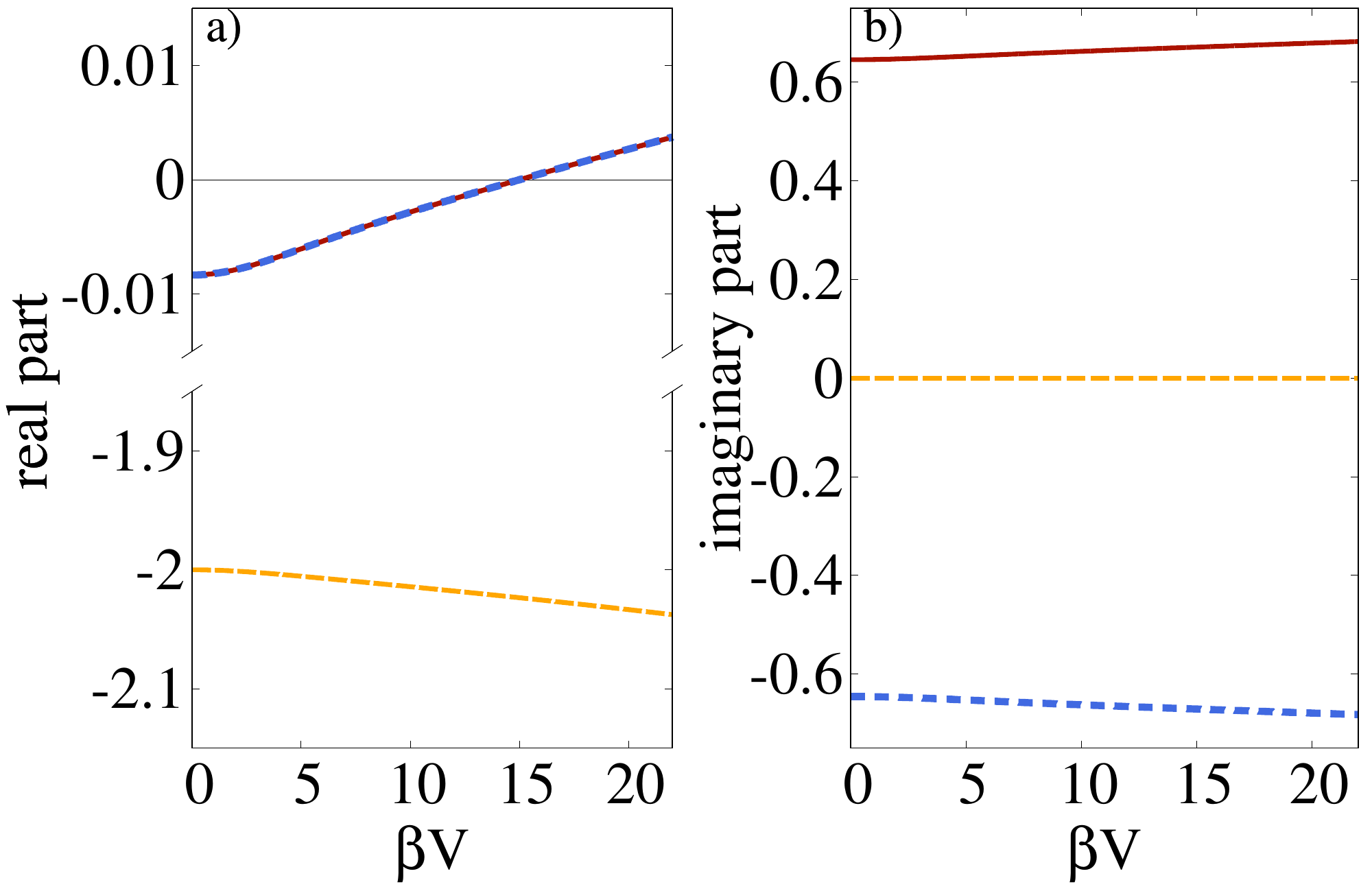}
\end{center}
\caption{Eigenvalues of $J_0$ evaluated at the steady state as a function of the applied bias voltage. As $V$ is increased the two complex conjugated eigenvalues (solid red and dotted blue) become purely imaginary at a critical value of $\beta \bar V_{\text{cr}} = 15$, which denotes the bifurcation point and the stability of the fixed point changes. Further increase of $V$ results in a stable limit cycle and an unstable fixed point.}
\label{fig:StabilityEV}
\end{figure}

The stability of the fixed point is determined by the eigenvalues of the Jacobian $J_0$ of the right hand side of the latter equation evaluated at steady state ($\dot{\bar x}=0$, $\dot{\bar v}=0$, $\dot{\bar p}_1=0$): If one or more eigenvalues have positive real part the fixed point is unstable. Fig.~\ref{fig:StabilityEV} shows the real part (left) and imaginary part (right) of the three eigenvalues as a function of the applied bias voltage. There exists a pair of conjugate eigenvalues (solid red and dotted blue) which become purely imaginary at a critical value $\beta \bar V_{\text{cr}} = 15.0$. At this point the Hopf bifurcation sets in and the MF system undergoes the transition from a stable fixed point to a stable limit cycle.

\end{document}